\documentclass[fleqn,usenatbib]{mnras}

\usepackage{newtxtext,newtxmath}

\usepackage[T1]{fontenc}

\DeclareRobustCommand{\VAN}[3]{#2}
\let\VANthebibliography\thebibliography
\def\thebibliography{\DeclareRobustCommand{\VAN}[3]{##3}\VANthebibliography}

\usepackage{graphicx}	
\usepackage{amsmath}	
\usepackage{booktabs}   
\usepackage{multirow}   
\usepackage{array}
\usepackage{caption}
\usepackage{comment}
\usepackage[mathlines]{lineno}

\newcommand{\msun}{\mbox{${\rm M}_{\odot}$}}
\newcommand{\Nion}{$\langle N_{\rm ion} \rangle~$}

\newcommand{\Novii}{$\langle N_{\rm O ~ VII} \rangle~$}
\newcommand{\Noviii}{$\langle N_{\rm O ~ VIII} \rangle~$}
\newcommand{\fion}{$f_{\rm ion}~$}
\newcommand{\fovi}{$f_{\rm O ~ VI}~$}
\newcommand{\fovii}{$f_{\rm O ~ VII}~$}
\newcommand{\foviii}{$f_{\rm O ~ VIII}~$}

\defcitealias{Faerman2020}{FSM20}
\defcitealias{Faerman2023_Mcool}{FW23}

\title[Oxygen in the CGM]{Highly Ionized Oxygen in the Circumgalactic Medium of TNG100 Galaxies: Distributions, Thermal States, and Ionization Mechanisms}

\author[Oren et al.]{
Yossi Oren,$^{1}$\thanks{E-mail: orenyossi01@gmail.com}
Amiel Sternberg,$^{2, 1, 3}$
Christopher F. McKee,$^{4, 5}$ and
Yakov Faerman $^{1}$ 
\\  
$^{1}$School of Physics and Astronomy, Tel Aviv University, Ramat Aviv 69978, Israel\\
$^{2}$Center for Computational Astrophysics, Flatiron Institute, 162 5th Ave., New York, NY, 10010, USA \\
$^{3}$Max-Planck-Institut f\"ur extraterrestrische Physik (MPE), Giessenbachstr., 85748 Garching, Germany \\
$^{4}$ Department of Physics, University of California, Berkeley, CA 94720, USA \\
$^{5}$ Department of Astronomy, University of California, Berkeley, CA 94720, USA
}

\date{Accepted XXX. Received YYY; in original form ZZZ}

\pubyear{2026}

\begin{document}
\label{firstpage}
\pagerange{\pageref{firstpage}--\pageref{lastpage}}
\maketitle

\begin{abstract}
We investigate the origin of the highly ionized oxygen species, O VI, O VII, and O VIII, in the circumgalactic medium (CGM) of galaxies at redshifts $z \approx 0$, by post-processing the outputs of the Illustris TNG100 simulation alongside analytic modeling. Our computations are broadly consistent with observations, and we provide convenient functional fits for the predicted mean oxygen ion column densities as functions of halo virial mass. We determine the relative roles of electron impact collisional ionization versus photoionization by the metagalactic radiation field in producing the ions for halo masses ranging from $10^{10.5}$ to $10^{14.5}$~M$_\odot$. We also determine the thermal gas phases, virialized hot, intermediate temperature, or cool, within which the ions reside. If the halo virial temperature is below the gas temperature at which the ion fractions are maximal for collisional ionization equilibrium (CIE), then photoionization dominates the overall CGM ion mass and column density. When photoionization dominates, the ions are produced primarily in the shock heated virialized phase. If the virial temperature is above the gas temperatures at which the ion fractions are maximal for CIE, then collisions dominate the ionization. However, if the virial temperature is much larger than the ion CIE peak temperature a substantial portion of the ion mass may reside within the intermediate temperature phase as opposed to the hot virialized gas. For MW-mass halos, the O VI and O VII are produced collisionally, and the O VIII by photoionization, all three in the hot, virialized phase.

\end{abstract}

\begin{keywords}
galaxies: evolution -- galaxies: formation -- galaxies: halos
\end{keywords}



\section{Introduction} \label{sec:intro}
The circumgalactic medium (CGM) is known to play a central role in galaxy evolution, acting as both the source of accreting material and the repository of galactic feedback \citep{Somerville2015_CGM_intro, Tumlinson2017_CGM_intro, Naab2017_CGM_intro, Oren2026}. Recent observations and simulations have established the CGM as multiphase and dynamically complex, spanning a wide range of gas densities, temperatures, and ionization states. Oxygen, being the most abundant metal in the universe, is a particularly useful probe of the physical conditions of this gas, as it traces both enrichment processes and thermodynamic structure.

Among oxygen ions, the highly ionized states O VI, O VII, and O VIII are especially important probes of the warm-hot CGM \citep{Bregman2007_O_ions, Tumlinson2011_COS_halos, Faerman2017}. These ions may arise in hot gas with characteristic temperatures of $T \sim 10^{5.5}-10^{6.5}$ K due to collisional ionization, or in diffuse gas with number densities $n \lesssim 10^{-4.5} ~\rm{cm^{-3}}$ due to photoionization by the metagalactic ionizing background \citep[][hereafter \citetalias{Faerman2020}]{Faerman2020}. The oxygen ions trace distinct spatial locations and thermal phases of the CGM, from cooling interfaces and galactic outflows, to shock heated virialized hot halos out to the intergalactic medium. 

Observationally, O VI has been extensively studied through ultraviolet absorption-line (${\rm \lambda\lambda}$ 1031.9 1037.6 \AA) surveys in quasar sightlines such as COS-Halos \citep{Tumlinson2011_COS_halos}, CUBS \citep{Chen2020_CUBS}, and CGM\textsuperscript{2} \citep{Tchernyshyov2022_obs}. These have revealed ubiquitous, massive oxygen-rich halos around star-forming galaxies extending to distances comparable to the halo virial radii. These observations demonstrate that a substantial fraction of galactic metals reside in the CGM, often exceeding the metal mass within the central galaxies themselves \citep{Peeoples2014,Nishigaki2025}.

The higher ionization states O VII and O VIII are typically accessible through X-ray spectroscopy (21.6 \AA~$K_\alpha$, and 18.6 \AA~$K_\beta$ for O VII, and 18.97 \AA~$K_\alpha$ for O VIII) which remains observationally challenging. Existing O VII measurements, largely limited to the Milky Way \citep{Bregman2007_OVII_MW, Gupta2012_OVII_MW, Miller2013_MW_OVII} and a small number of nearby systems \citep{Nicastro2023_OVII_abs, Mathur2023_OVII_abs}, suggest that these ions trace a hot, relatively inner component of the CGM. O VIII lines have so far only been measured from within the Milky Way \citep{Das2019_OVIII_MW, Locatelli2024_OVIII_MW}, yet given the MW-only detections it is unclear where the gas resides. 

A key complication in interpreting the measured oxygen column densities is the set of assumptions required to infer physical quantities such as gas mass, density, and spatial extent. In most observational analyses, the conversion from O VI column density to a total oxygen (or gas) mass depends sensitively on the assumed ionization mechanism and ionization fraction. A common first approach is to assume collisional ionization equilibrium (CIE) in hot gas \citep[see e.g.][]{Tumlinson2011_COS_halos, Gupta2012_OVII_MW, Das2019_OVIII_MW}. For example, in CIE the O VI fraction peaks at $T \sim 10^{5.5}$ K. But O VI can also be produced by photoionization, 
which is often understood to occur in cooler gas, $\sim 10^4$~K, that is in thermal equilibrium with the ionizing radiation field \citep[e.g., ][]{Stern2016_cool_OVI}. Naturally, the inferred properties of the CGM, such as its gas mass, will vary significantly depending on the ionization mechanisms at play. As a result, distinguishing between photoionized and collisionally ionized origins of O VI is not only a question of physical interpretation but also directly impacts quantitative conclusions drawn from observations.

Complementary to ionization considerations, the physical distribution of O VI within the CGM provides important insight into the structure and dynamics of halo gas. Observationally, O VI absorption is detected out to large impact parameters, suggesting extended distribution out to the halo virial radii or beyond \citep{Tumlinson2011_COS_halos, Faerman2017, Stern2018_ovi}, but the underlying three-dimensional structure remains poorly constrained. Previous simulation analyses have shown that O VI is more radially extended than cooler CGM tracers and occupies thermal phases distinct from the hot gas traced by O VII and O VIII \citep[e.g. ][]{Oppenheimer2016_ovi_EAGLE, Nelson2018_TNG_absorption}. 

Cosmological hydrodynamical simulations provide a crucial framework for interpreting these observations as they allow, within their limitations, detailed analyses of individual galactic and halo components in three dimensions. 
The IllustrisTNG simulations \citep{Marinacci2018_TNGintro, Springel2018, Nelson2018, Naiman2018, Pillepich2018_results} that we use in this work have enabled systematic studies of highly ionized oxygen across a wide range of halo masses and environments. Particularly significant work is the \cite{Nelson2018_TNG_absorption} analysis of O VI, O VII, and O VIII in Illustris TNG100 and TNG300, in a post-processing study. \cite{Nelson2018_TNG_absorption} find good agreement between observed and simulated O VI column densities and column density distribution functions (CDDFs). They provide number and column density profiles for O VI out to very large distances, and decomposed into 1-halo and 2-halo terms. In \cite{Nelson2018_TNG_absorption}, the masses and mean column densities of O VI, O VII and O VII are presented as functions of galaxy and halo mass, and their CDDFs are also generated and evolved with redshift. 

However, several fundamental questions remain unresolved. First, which ionization process (collisional vs photoionization) dominates the O VI, O VII, and O VIII production? Second, which gas phase — in terms of temperature, density, and dynamical state — host these ions? Third, how are the ions spatially distributed within halos, and what are the characteristic half-mass scale radii and volume filling fractions? In this work, we address these questions by combining outputs of the IllustrisTNG100 simulation with post-processing computations of the ion abundances, together with analytic modeling and comparisons to observational constraints. 

In \S~\ref{sec:ionization}, we briefly review the roles of collisional ionization and photoionization for the ion production. In \S~\ref{sec:illustris} we describe our TNG100 halo sample. In \S~\ref{sec:oxymaps} we present results of our oxygen ion abundance computations in the form of maps and column density profiles, which we compare to observations. In \S~\ref{sec:plm} we make use of our analytic ``power-law model'' to represent and analyze the TNG100 results. In \S~\ref{sec:mean_columns} we present computations of the mean halo column densities as functions of halo mass, and discuss analytic forms for the mean columns. We use these results to identify the dominant ionization mechanisms, collisional versus radiative, for a given halo mass. In \S~\ref{sec:ions_different_mvirs} we present phase diagrams for three representative halo scales, low-mass, Milky Way mass, and high-mass, to illustrate the distinct gas phases, including cooling properties, within which the oxygen ions are produced. We summarize in \S~\ref{sec:summary}.

\section{Ionization} \label{sec:ionization}
Our focus is on the competition between two primary ionization mechanisms, electron impact collisional ionization, and photoionization by the metagalactic radiation field (MGRF). We do not consider additional sources of radiation such as active galactic nuclei (AGN), X-ray binaries, or large scale galactic winds  \citep{Oppenheimer2018_OVI_AGN, Upton_Sanderbeck2018_UVB, Sarkar2022}. These additional sources may compete with the MGRF in the inner parts of the CGM, but the higher gas densities at these locations mitigate their effects relative to collisional ionization.

\begin{figure}
        \includegraphics[width=0.45 \textwidth]{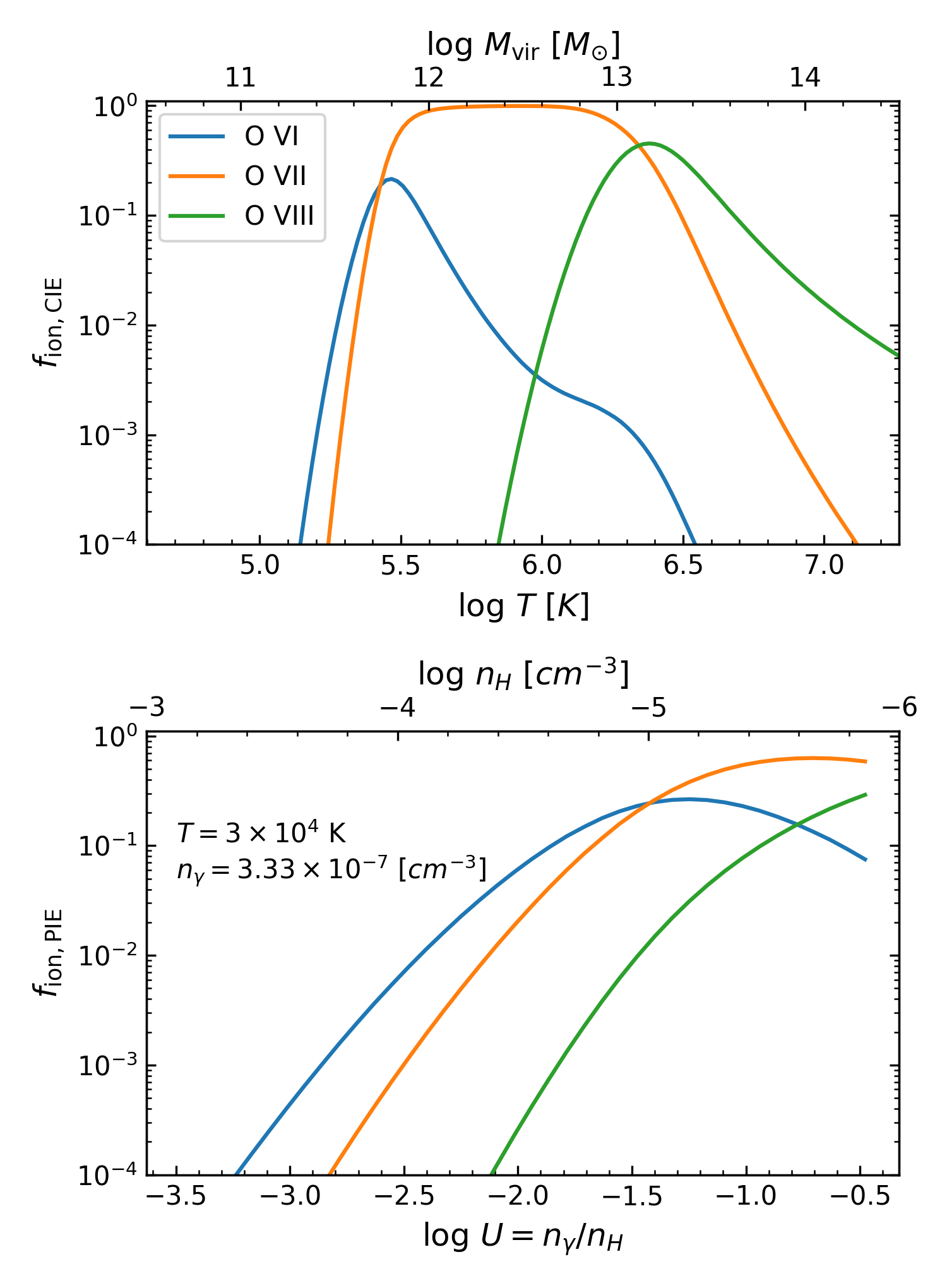}
        \caption{Top: Ion fractions assuming collisional ionization equilibrium (CIE) \citep{Gnat2007_CIE} for O VI (blue), O VII (orange), and O VIII (green) versus gas temperature $T$. The upper $x$-axis is halo virial mass, $M_{\rm vir}$, assuming $T_{\rm vir}=T$.
        Bottom: Ion fractions assuming photoionization equilibrium (PIE) in the $z=0$ metagalactic radiation field (MGRF) at $T=3\times 10^4$~K.}
        \label{fig:f_ion_CIE}
\end{figure}

The efficiency of collisional ionization depends primarily on the gas temperature which must be high enough to enable electron stripping to O VI, O VII, and O VIII. For these ions the required cumulative ionization energies are 294.98, 433.10, and 1172.39 eV respectively. In collisional ionization equilibrium (CIE), the ion populations are in a formation-destruction steady state and do not vary with time. The ionization balance is determined by the competition between collisional ionization and radiative and dielectroninc recombination, leading to ion fractions that depend exclusively on the gas temperature, independent of gas density. In Fig. \ref{fig:f_ion_CIE} (upper panel) we plot the ion fractions, $f_{\rm ion} \equiv n_{\rm ion} / n_{\rm O}$, where $n_{\rm ion}$ is the oxygen ion density and $n_{\rm O}$ is the total oxygen density, for O VI (blue line), O VII (orange line), and O VIII (green line) as a function of gas temperature, $T$, taken from \cite{Gnat2007_CIE}. 

The CIE ion fractions can also be presented as functions of halo mass, $M_{\rm vir}$ (upper $x$-axis), by setting $T_{\rm vir}=T$ where $T_{\rm vir}$ is the halo virial temperature (see \S~\ref{sec:defs} below for our definitions of $M_{\rm vir}$ and $T_{\rm vir}$). The ion fractions peak at specific gas temperatures and virial masses, and we shall henceforth refer to these maxima as the ``CIE peaks'', and the temperatures at which they occur as the ``CIE peak temperatures''. Fig. \ref{fig:f_ion_CIE} provides intuition into the typical halo masses where each ion is expected to peak under CIE; O VI at $T \approx 3\times 10^{5}$ K, and $M_{\rm vir} \approx 5\times10^{11} ~ \rm{M_{\odot}}$, with \fovi $\approx 0.2$; O VIII at $T \approx 3\times 10^{6}$ K, and  $M_{\rm vir} \approx 2\times10^{13} ~ \rm{M_{\odot}}$, with \foviii$\approx 0.5$; and O VII with \fovii $\approx 1$ for an extended temperature range between the O VI and O VIII peaks. The rise of each ion fraction is exponentially steep as a function of temperature: $\sim 3$ dex rise in \fion for a rise of $\sim 0.25$ dex rise in $T$. The decrease is less steep.

Second is photoionization. In the CGM, the dominant source of ionizing photons is generally assumed to be the metagalactic radiation field (MGRF) produced by background quasars and star-forming galaxies, although local radiation sources may also contribute close enough to the central galaxies \citep{Upton_Sanderbeck2018_UVB, Faerman2023_Mcool}. The efficiency of photoionization depends primarily on the ionization parameter, $U=n_\gamma/n_H$, the ratio of the local ionizing photon density to the gas density. As a result, highly ionized species can exist in relatively cool gas,  at temperatures well below the CIE peaks, provided the density is sufficiently low and the radiation field is sufficiently intense. In the lower panel of Fig.~\ref{fig:f_ion_CIE} we show the ion fractions, $f_{\rm ion,PIE}$, as functions of $U$, assuming photoionization equilibrium (PIE) at a gas temperature of $3\times 10^4$~K, in the presence of the optically thin \cite{Haardt_Madau2012} MGRF at $z=0$. The temperature of $3\times 10^4$~K is representative of gas in thermal equilibrium with the MGRF). For these computations we made use of the {\it Cloudy} photoionization code \citep{Ferland2017_CLOUDY}. For the assumed MGRF, $n_\gamma=3.33\times 10^{-7}$~cm$^{-3}$ above the Lyman limit (13.59 eV). The fractions of this MGRF photon density above the photoionization energies of 113.90, 138.12, and 739.29 eV, required for the formation of O VI, O VII, and O VIII, are 0.09, 0.07, and 0.02 respectively. The gas densities corresponding to our $n_{\gamma}$ (upper $x$-axis) span the density range expected in the warm-hot CGM of galaxies at redshift $z=0$.

\begin{figure*}
        \makebox[\textwidth][c]{\includegraphics[width= \textwidth] {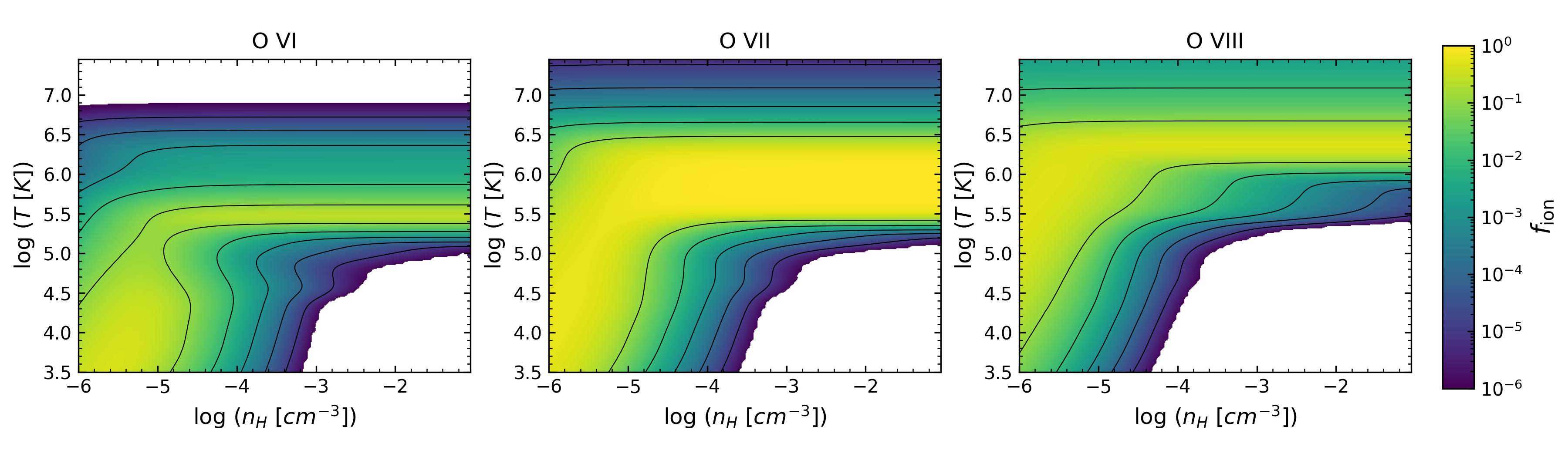}}
        \caption{Ion fractions for O VI (left panel), O VII (central panel), and O VIII (right panel), calculated assuming collisional ionization by thermal electrons together with photoionization by the \citet{Haardt_Madau2012} metagalactic radiation field at redshift $z=0$, as a function of hydrogen density, $n_H$, and gas temperature, $T$. Contours mark ion fractions in 1 dex increments between $10^{-5}$ and $10^{-1}$.}
        \label{fig:f_ion_full}
\end{figure*}

We have also carried out {\it Cloudy} computations of the
ion fractions assuming the combination of collisional ionization by thermal electrons and MGRF photoioinzation, for any input hydrogen gas density, $n_H$ and temperature $T$.  In Fig. \ref{fig:f_ion_full} we show the ion fractions for O VI (left panel), O VII (middle panel), and O VIII (right panel) as functions $n_H$ and $T$ \citep[see also Fig.~3 of][]{Nelson2018_TNG_absorption}. 
For each ion, there are two regions in the $(n_H, T)$ plane where the fractions are maximal. First are near the CIE peak temperatures, largely independent of density (the horizontal yellow strips), and for which the peak fractions are as shown in Fig. \ref{fig:f_ion_CIE} upper panel. Second, are at relatively low densities and temperatures where photoionization dominates as shown in Fig. \ref{fig:f_ion_CIE} lower panel. For O VI this is at $-6 \lesssim \log(n_H / cm^{-3}) \lesssim -5$ and $3.5 \lesssim \log(T/K) \lesssim 4.5$; for O VII at $\log(n_H / cm^{-3}) \lesssim -5$ and $3.5 \lesssim \log(T/K) \lesssim 5.5$; and for O VIII at $\log(n_H / cm^{-3}) \lesssim -5.5$ and $4.5 \lesssim \log(T/K) \lesssim 6.5$. In the photoionization limit, the maximal O VI ionization fraction is 0.4, a factor 2 larger than for CIE. However, the maxima for O VII and O VIII are larger for CIE compared to PIE.

As we calculate ionization states and other related quantities in this work, we will label them as ``CI+PI'' when assuming both collisional ionization by thermal electrons and photoionization by the MGRF, and ``CIE only'' when assuming no photoionization and only collisional ionization. We emphasize that CI+PI and CIE are alternate steady-state solutions for the ionization states, the first depending on $n_H$ and $T$, the second on $T$ only.

\section{I\MakeLowercase{llustris} TNG100}\label{sec:illustris}
The IllustrisTNG (The Next Generation) simulation is a state-of-the-art cosmological hydrodynamical simulation developed to investigate galaxy formation, the evolution of large-scale structures, and the interplay between dark matter and baryonic matter \citep{Marinacci2018_TNGintro, Springel2018, Nelson2018, Naiman2018, Pillepich2018_results}. The simulation employs the \textsc{AREPO} code \citep{Springel2010}, which uses a moving-mesh finite-volume approach to solve the coupled equations of magneto-hydrodynamics (MHD), together with a tree-based method to calculate gravitational interactions \citep{Pakmor2011, Pakmor2013}.

Physical processes at scales smaller than the resolution of the IllustrisTNG simulation are accounted for using subgrid models. Star formation is described by a density-dependent prescription in which sufficiently dense and cool gas can form stars; a chemical enrichment model follows the production and evolution of elements in stars and their subsequent distribution into the surrounding gas; stellar feedback is implemented primarily through supernovae that inject energy into their surrounding medium and drive galactic winds; radiative processes account for gas heating due to the universal background radiation and local AGN radiation, alongside gas cooling via metal-lines; and supermassive black holes are seeded and subsequently grow, releasing energy into their surroundings through both thermal and kinetic modes of AGN feedback \citep{Pillepich2018, Weinberger2017}.

The cooling function in IllustrisTNG accounts for the ionization states of the individual elements using precomputed {\it Cloudy} tables, but the ionization states are not saved as simulation outputs \citep{Vogelsberger2013_sim_model}. In our work we recompute the ionization states in post-processing, self-consistently with the simulation particle data, including gas density, temperature, and the background MGRF.

The IllustrisTNG simulation suite consists of three box sizes, with side lengths of 51.7, 110.7, and 302.6 Mpc, each simulated at three different resolutions, designated from 1 (highest resolution) to 3 (lowest resolution). In this work, we use the TNG100-1 simulation, which provides the highest resolution for the 110.7 Mpc box. This simulation contains $1820^3$ dark matter particles, each with a mass of $7.5 \times 10^6 ~\msun$, as well as gas particles with a mass resolution of $1.4 \times 10^6 ~\msun$. Stars form with a comparable mass resolution. As dark-matter halos and galaxies form and grow throughout the simulation, they are identified using two complementary algorithms. The Friends-of-Friends (FoF) algorithm identifies dark-matter halos, while the \textsc{SUBFIND} algorithm \citep{Springel2001} identifies gravitationally bound substructures within these halos, corresponding to individual galaxies.

The cosmological parameters adopted by IllustrisTNG are based on the results of \cite{Planck2016}, with a Hubble constant of $H_0 = 67.74 ~{\rm km ~s^{-1} Mpc^{-1}}$, matter and baryon density parameters of $\Omega_{m, 0}=0.3089$ and $\Omega_{B, 0}=0.0486$, respectively, and a cosmological constant of $\Omega_{\Lambda, 0}=0.6911$. The simulation follows the evolution of the universe from high redshift ($z = 127$) to the present day ($z = 0$) across 100 snapshots. At $z=0$, more than $6 \times 10^6$ dark matter halos are identified, containing a total of more than $4 \times 10^6$ galaxies.

\subsection{TNG100 halo sample at \texorpdfstring{$z$=0}{z=0}} \label{sec:TNG_sample}
\begin{figure}
        \includegraphics[width=0.48 \textwidth]{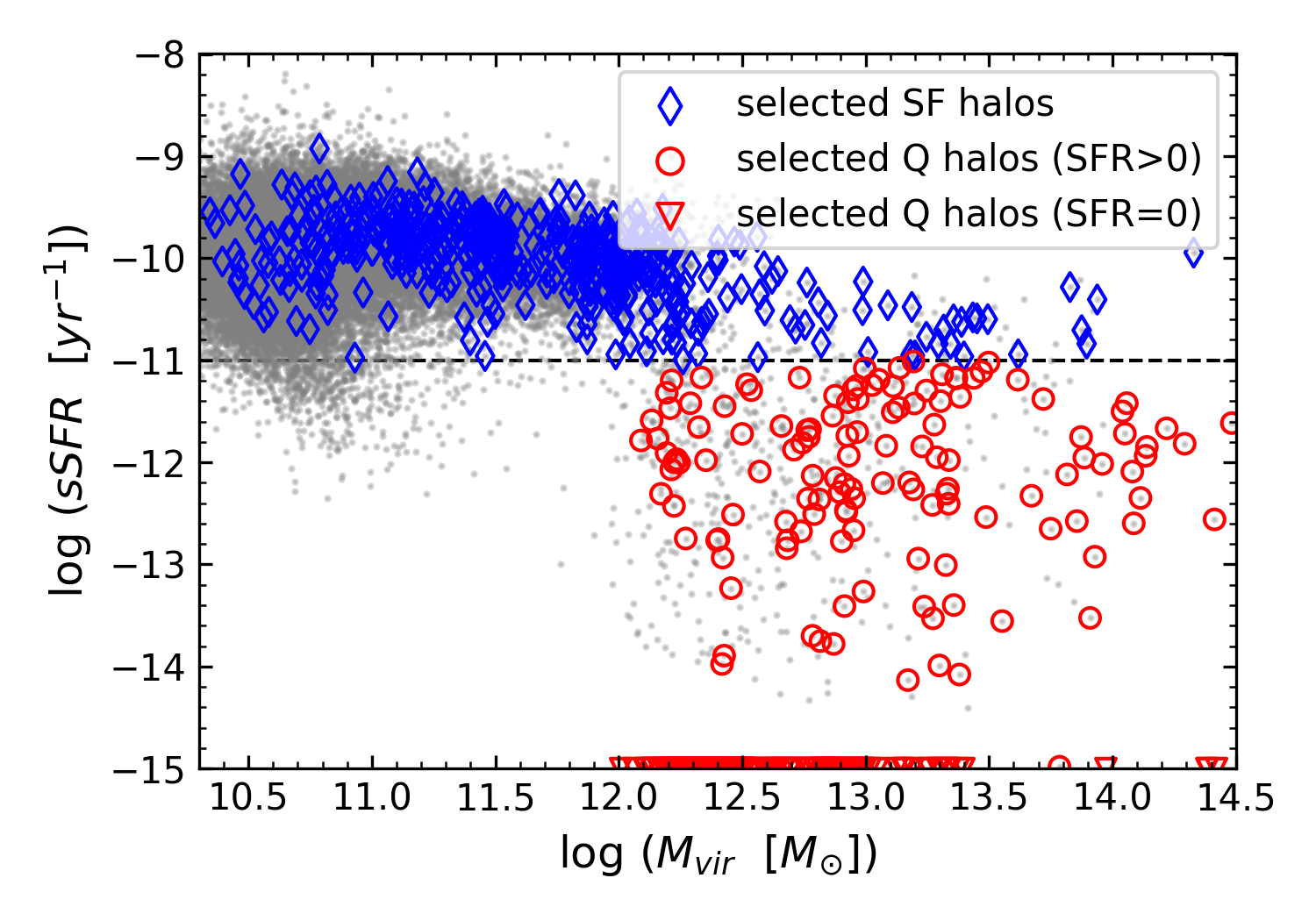}
        \caption{Our sample of 785 TNG100 halos selected for analysis in this work, presented on the specific star formation rate ($sSFR$) versus halo virial mass plane. Blue diamonds are halos with central star-forming galaxies. Red circles are halos with quenched galaxies, and red triangles are for galaxies with no star formation whatsoever. The black dashed line at $sSFR=10^{-11}$~yr$^{-1}$ is the dividing line between star-forming and quenched systems. The gray points are the full TNG100 halo population within this mass range.}
        \label{fig:sample_sSFR}
\end{figure}

Oxygen absorption in the CGM has been studied observationally in a diverse population of galaxies, spanning a large range of galaxy masses and star formation rates. Therefore, to compare with the observed populations, in this work we build a sample selected from $z=0$ TNG100 galaxies with halo masses ranging from $M_{\rm vir} = 10^{10.5}$ to $10^{14.5} ~ \rm{M_{\odot}}$. We include in our sample both star forming and quenched galaxies, and we define the limit for star formation as $sSFR = SFR / M_* > 10^{-11} ~ \rm{M_{\odot} ~yr^{-1}}$. We analyze the same halos selected in \cite{Oren2024}, in addition to 137 star forming galaxies with $M_{\rm vir}$ between $10^{10.5} ~\msun$ to $10^{11} ~\msun$. The \cite{Oren2024} sample consists of 150 star forming galaxies that trace the $sSFR$ main sequence of TNG100; 50 galaxies whose $sSFR$ is 0.5 dex above the main sequence; 50 galaxies that are 0.5 dex below it; 150 quenched galaxies with $10^{12} < M_{\rm vir} / \msun < 10^{14.5}$; and 248 extra randomly selected galaxies. In total, we analyze 785 galaxies in this work, 284 of which are quenched and 501 are star forming. We cover the full $sSFR$ range for the halo masses we sample from TNG100, as shown in Fig. \ref{fig:sample_sSFR}.

IllustrisTNG associates dark matter particles to halos using the Friends-of-Friends (FoF) algorithm, and finds substructures (i.e. galaxies) within the halos with the SUBFIND algorithm \citep{Springel2001}. However, we find that at larger distances from the central galaxies the FoF algorithm tends to miss particles that are physically near the halo and would contribute to observations. Therefore, when analyzing our TNG100 halo sample we select particles based on their distance from the center of the halo and not based on their association with one galaxy or another (we perform several cuts on which we will elaborate later). However,galaxy and halo parameters given by the simulations (e.g. halo virial mass, stellar mass, star formation rates) are calculated for particles based on their association with a FoF halo or a SUBFIND galaxy. 

By extending our sample to lower halo masses, we should consider the possibility that we encounter unresolved galaxies that are prone to shot noise. The signal-to-noise ratio for particle based results depends on the square root of the number of particles, and it is common to consider resolved galaxies as those that have more than $100$ particles such that relative error is smaller than $10 \%$ \citep[as is done by e.g. ][]{Pillepich2018}. Considering the mass resolution of TNG100, this translates to a stellar mass of $\sim 10^{8} ~\msun$, equivalent to a halo mass of $\sim 10^{10.8} ~\msun$ in TNG100. Therefore, we expect results from our new set of halos, all of which have $M_{\rm vir} < 10^{11} ~\msun$ to be more prone to numerical errors (and particularly their galactic properties).

\subsection{Definitions and shorthands} \label{sec:defs}

In this work we define the halo virial mass, $M_{\rm vir}$, as the mass of a virialized structure whose mean matter density is a factor $\Delta_c\approx 100$ times the cosmic critical density $\rho_c = 8.53\times 10^{-30}$~g~cm$^{-3}$ at $z=0$ \citep{Bryan1998}. The associated virial radius is then given by
\begin{equation} \label{eq:R_vir}
\begin{split}
    R_{\rm vir} \  & \equiv \  \bigl(\frac{4\pi}{3} \Delta_c \rho_c\bigr)^{-1/3}M_{\rm vir}^{1/3} \\
    & = \ 267 \ \Bigl(\frac{M_{\rm vir}}{10^{12} ~ \msun}\Bigr)^{1/3} E^{-2/3}(z) \ {\rm kpc} \ \ \ .
\end{split}
\end{equation}
The factor $E^2(z) \equiv \rho_c / \rho_{c, 0}= \Omega_m (1+z)^3 + \Omega_{\Lambda}$ accounts for the increase in the critical density with redshift. The virial mass is the total mass within the virial radius, and both quantities are output by the simulation. Following \citet{Faerman2017} and \citet{Oren2024} we define the virial temperature as
\begin{equation} 
\begin{split}
    T_{\rm vir} & \ \equiv \ \frac{1}{3}\left(\frac{G {\bar m}}{k_{\rm B}}\right)\frac{M_{\rm vir}}{R_{\rm vir}},\\
     & =3.90\times 10^5\Bigl(\frac{M_{\rm vir}}{10^{12}\ \msun}\Bigr)^{2/3} E^{2/3}(z)\ \ {\rm K} ~~,
\end{split}
\end{equation}
where $\bar{m} \approx 0.6 m_p$ is the mean mass per gas particle, $k_{\rm B}$ is the Boltzmann constant, and $G$ is the gravitational constant\footnote{In some conventions $T_{\rm vir}$ is calculated with a coefficient of $1/2$ rather than $1/3$, but TNG100 halos appear to agree with the latter (see Fig. 11 of \citealp{Oren2024}).}. 

The radius of the central galaxy is defined as twice the stellar-half-mass radius, and the simulation outputs the radius together with the enclosed stellar mass and star formation rate. For our analysis we define CGM as gas particles that are not star forming (i.e. have $SFR = 0$), and are outside twice the stellar-half-mass radius of any galaxy, be it central or satellite.

\section{Oxygen column density maps in TNG100}\label{sec:oxymaps}

\subsection{Generating the maps}\label{subsec:n_ion_maps}
We start our analysis of the ion populations in TNG100 by constructing column density maps of O VI, O VII, and O VIII, in post-processing. We set the $xy$ size of any map equal to ${\rm max}(800~{\rm kpc}, 2.2R_{\rm vir})$. We then integrate along the lines of sight (the $z$ direction) in two ways. First, we include only particles within the virial radius. We refer to these as "halo cut" maps. Second, we isolate all gas particles within 2.1 Mpc radius spheres around the halo centers, and integrate through the spheres along lines of sight contained within the full $xy$ map dimensions. We dub these as "2Mpc cut" maps. The 2.1~Mpc length scale is motivated by the velocity analysis of O VI absorbers by \cite{Tumlinson2011_COS_halos}, showing that most O VI absorbers in the COS-Halos survey are within $150$ km $\rm s^{-1}$ with respect to the galaxy redshift, equivalent to a Hubble flow size of 2.1 Mpc. 

We note that the velocity cut adopted for O VI may not be appropriate for O VII and O VIII, as recent observations suggest possible velocity variations from the galaxy far beyond $150 ~\rm{km ~s^{-1}}$. The small sample of external O VII absorption lines alongside the rough spectral resolution used by modern instruments do not allow a proper limit on the velocity distribution. For example in XRISM, the telescope used by \citealp{Nicastro2023_OVII_abs}, the spectral resolution is $\sim 5$ eV, corresponding to a velocity width of $\sim 3000 ~\rm{km~s^{-1}}$. Additionally, observations of O VII and O VIII from within the MW may be affected by gas anywhere between the galactic disk and the local Group. We therefore opt to discuss maps from the 2Mpc cut for O VII and O VIII as well.

The maps are constructed only for gas particles we consider CGM (i.e. are not star forming and are outside of any galaxy). For each gas particle we calculate the volume density of each oxygen ion using the following relation:
\begin{equation} \label{eq:n_ion}
    n_{\rm ion} = n_H x_{\rm O} ~f_{\rm ion}(n_H, T)
\end{equation}
where $n_H$ is the hydrogen number density (cm$^{-3}$), and  $x_{\rm O} = n_{\rm O} / n_H$ is the oxygen abundance relative to hydrogen by number, and $f_{\rm ion}(n_H,T)$ is the ion fraction (assuming either CI+PI or CIE). IllustrisTNG calculates the oxygen \textit{mass} fractions $m_{\rm O}/m_{\rm gas}$ relative to the total gas mass of the particle, which we convert to the number fraction via 
\begin{equation}\label{eq:f_element}
    x_{\rm O} = \left(\frac{m_{\rm O}}{m_{\rm gas}} \right) / A_{\rm O} \left( \frac{m_{\rm H}}{m_{\rm gas}} \right) ~~ .
\end{equation}
where $A_{\rm O} \approx 16$ is the atomic weight of oxygen. 

With the ion volume densities in hand, we calculate the column densities using:
\begin{equation}\label{eq:N_integral}
    N_{\rm ion} = \int_{z_{\rm in}}^{z_{\rm out}} n_{\rm ion} dz 
\end{equation}
where $z$ is a coordinate along the line of sight, ranging from $z_{\rm in}$ to $z_{\rm out}$, determined by the particle cut in use. For a given impact parameter $b$ at the 2Mpc cut $z_{\rm out / in} = \pm \sqrt{(2.1 ~\mathrm{Mpc})^2 - b^2}$, and for the halo cut $z_{\rm out / in} = \pm \sqrt{R_{\rm vir}^2 - b^2}$. Because  the maps are bordered at the larger of 800 kpc or $2.2 R_{\rm vir}$, they have a spatial resolution of $\sim 1.56$ kpc per pixel. The maps are therefore at least $512 \times 512$ pixels, or $1068 \times 1068$ pixels for halos with $M_{\rm vir} = 10^{13.5} ~\msun$ whose $R_{\rm vir} \approx 833$ kpc. They are aligned with the arbitrary $(x, y)$ plane of the simulation such that the integral is through the $z$-axis of the simulation. As we sum over individual CGM particles, Eq. \ref{eq:N_integral} at a given pixel becomes:
\begin{equation}
    N_{\rm ion}(x,y) = \sum_{\rm i} n_{\rm ion,i} dl_{\rm i} ~~, 
\end{equation}
where $i$ is an index iterating over all CGM particles in either the halo cut or 2Mpc cut that intersect the $(x, y)$ coordinates of the pixel, and $dl$ is the path length within each gas particle. 

In Appendix \ref{appendix:maps} we show examples O VI, O VII, and O VIII column density maps, for both the 2Mpc and halo cuts, for the three specific TNG100 halos we discuss in detail in \S~\ref{sec:ions_different_mvirs}. 

\subsection{Comparison with O VI observations}
For a more robust analysis, we calculate azimuthaly-averaged profiles of our TNG100 column density maps as a function of $b / R_{\rm vir}$, the impact parameter normalized by the virial radius, out to $b = R_{\rm vir}.$

\begin{figure*}
        \makebox[\textwidth][c]{\includegraphics[width= \textwidth] {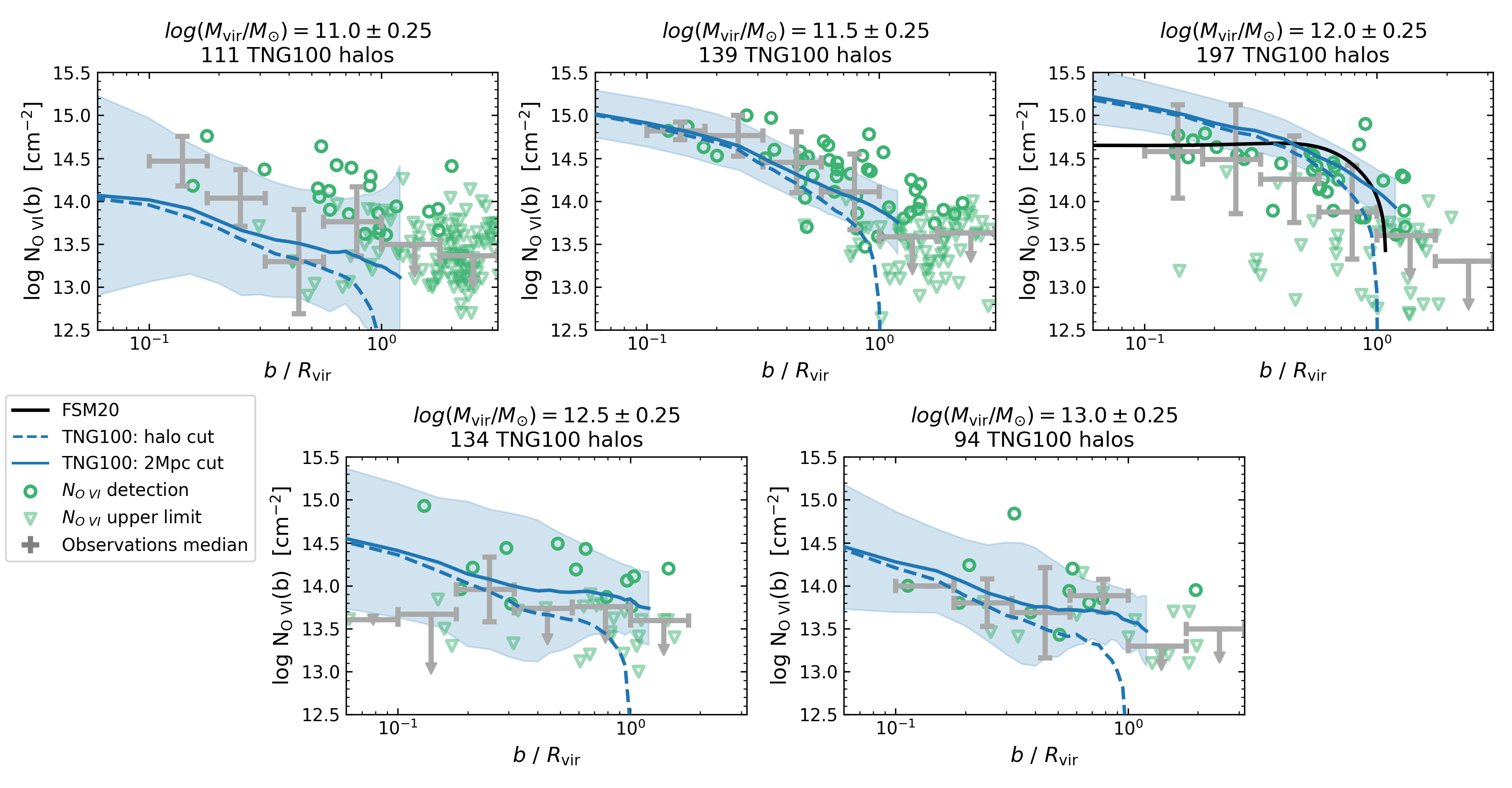}}
        \caption{Observations of O VI column densities presented against the radially averaged O VI column densities from TNG100, in half-dex mass bins centered around $\log(M_{\rm vir} / M_{\odot}) \in [11, 11.5, 12, 12.5, 13]$ (increasing from left to right, top to bottom). For  TNG100, profiles are taken from maps of the 2Mpc cut (solid lines surrounded by shaded regions marking the 16-84 percentiles) and halo cat (dashed lines). Observations are shown as empty green circles for detections and faded green triangles for upper limits. Gray errorbars represent the median observed $N_{\rm O~VI}$ in $b / R_{\rm vir}$ bins of 0.25 dex. Median observations are marked with downward facing arrows where the covering fraction in the bin is below 0.33. The \citetalias{Faerman2020} result for O VI is shown as a black solid line in the $10^{12} ~\msun$ panel.}
        \label{fig:ovi_observations}
    \end{figure*}

In Figure \ref{fig:ovi_observations} we compare the resulting TNG100 O VI column density profiles with observations obtained from three separate studies. First is\cite{Tchernyshyov2022_obs}, which aggregate O VI column density measurements from the CGM\textsuperscript{2} survey and other earlier studies. Their results include both star forming and quenched galaxies spanning redshifts of $0.0 < z < 0.6$, a stellar mass range of $M_* = 10^{7.8}-10^{11.2} ~\msun$, and impact parameters of up to 400 kpc. Second is from the CUBS survey and published by \cite{Qu2024}, showing galaxies with $0.43 < z < 0.72$, spanning a stellar mass range of $M_* = 10^{7.7}-10^{11.4} ~\msun$ and impact parameters of up to 1 Mpc. Third is also from the CUBS survey and published by \cite{Mishra2024}, focusing on star forming field dwarfs with $M_* = 10^{6.8}-10^{9.0} ~\msun$ at $0.07 < z < 0.73$ and with impact parameters up to 300 kpc. We convert all reported stellar masses to halo virial masses using the redshift dependent stellar to halo mass relation published by \cite{Moster2010} and perform our comparison for 5 mass bins, from $\log(M_{\rm vir} / M_\odot) = 11$ to $\log(M_{\rm vir} / M_\odot) = 13$ in 0.5 dex bins. The reported impact parameters are normalized by the virial radii associated with our computed virial masses (Eq. \ref{eq:R_vir})\footnote{The impact parameters from all surveys span a range of $0.05 < b/R_{\rm vir} < 4.5$, and the largest impact parameters are for the most massive galaxies.}. We mark positive detections by green circles, and upper limits by light-green triangles. We also calculate the median and standard deviations of the observed $N_{\rm O~VI}$ in $b/R_{\rm vir}$ bins of $0.25$ dex, shown as gray error bars. Standard deviations are replaced by downward facing arrows in impact parameter bins where the covering fraction is below $0.33$ (i.e. where over two thirds of observations in the bin are upper limits). The TNG100 2Mpc cut profiles are shown in solid lines, and the 16-84 percentiles are shaded in light blue. The halo cut profiles are shown in dashed blue lines. For the $10^{12}$~M$_\odot$ mass bin, the \citetalias{Faerman2020} result for the O VI profile is shown as a black solid line.

Because they include more gas, the 2Mpc cut profiles naturally produce higher ion column densities compared to profiles made with the halo cuts. Some of the differences may be due to background contributions from the intergalactic medium unassociated with the individual halos \citep{Stern2018_ovi, Ho2021_IGM_OVI, Bromberg2025_IGM_OVI}. We estimate the O VI IGM background contribution in TNG100 to be $10^{12.6}$~cm$^{-2}$ for a sight line of 4 Mpc based on the results by \cite{Nelson2018_TNG_absorption} (see their Fig.~9). However, the differences between our 2Mpc cut maps and our halo cut maps are generally larger than the background contribution. For example, for $M_{\rm vir}$ between 10$^{11}$ and 10$^{13}$~M$_\odot$, the differences range from $\sim 10^{13}$ to $10^{14}$~cm$^{-2}$, for comparable path lengths of $\sim 4$~Mpc. This suggests that for the lowest mass halos the background does contribute to the 2Mpc cut maps. At higher halo masses the background is negligible, and the excess columns are due to halo gas that extends beyond the nominal virial radii.

Comparing to observations, we note that all observations of O VI are consistent with a significant downturn in the number of detections at $b > R_{\rm vir}$. We find a reasonable agreement between the observed medians and the 2Mpc cut median within the virial radius for most mass bins. The observed medians are within the shaded regions (representing the 16-84 percentiles of the 2Mpc cuts) for the $10^{11} ~\msun$, $10^{11.5} ~\msun$, $10^{12.5} ~\msun$, and $10^{13} ~\msun$ mass bins. Of note is the $10^{12} ~\msun$ mass bin, for which our TNG100 results overestimate observations. While we cannot observationally constrain contributions from within the halos, the agreement between the 2Mpc cut and observations leads us to conclude that our TNG100 computations correctly reproduce the O VI population within the virial radii for a wide range of galaxy masses.


\subsection{Comparisons with O VII and O VIII observations}
\begin{figure}
        \includegraphics[width=0.45 \textwidth]{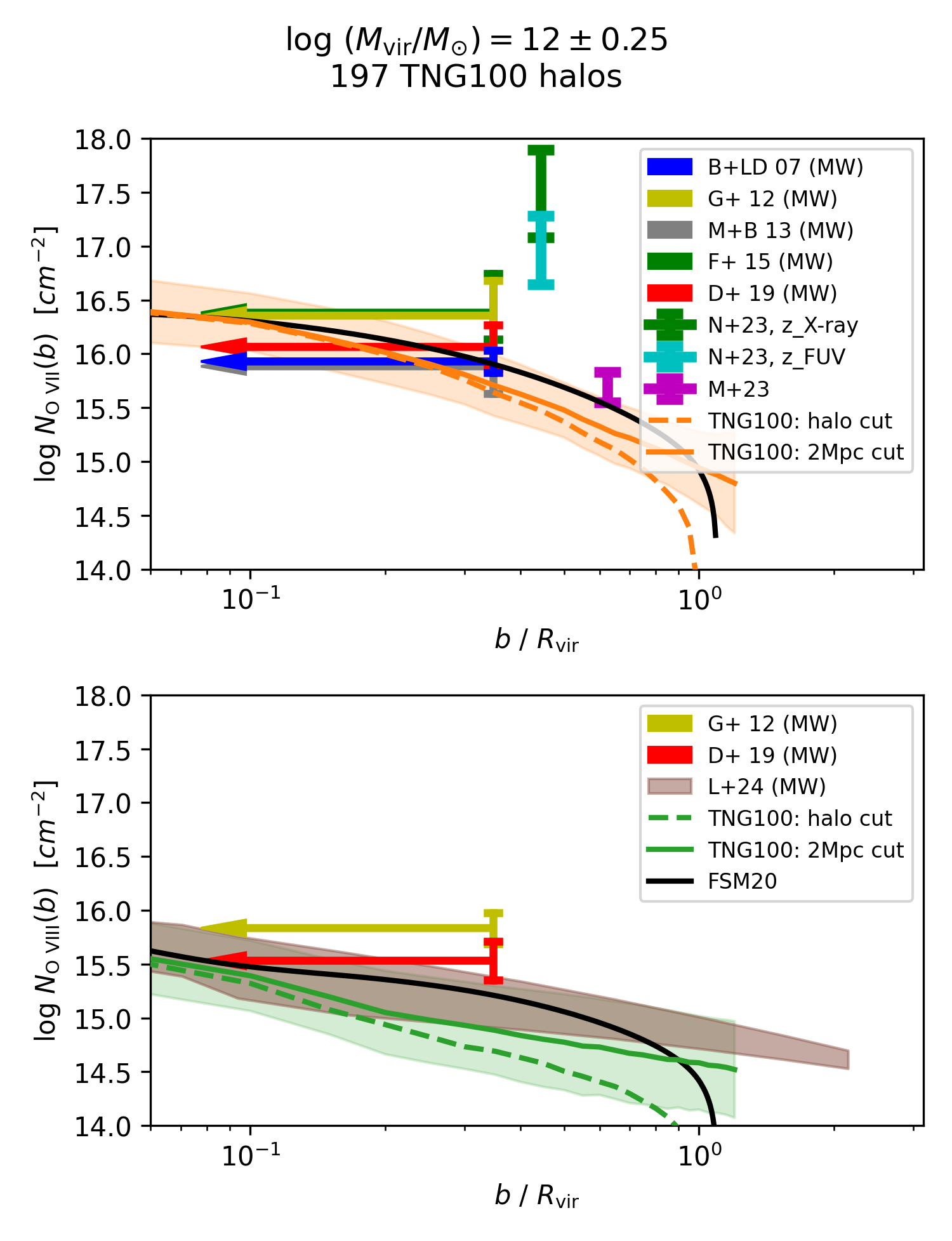}
        \caption{Observations of O VII (top panel) and O VIII (bottom panel) presented against the radially averaged column densities from TNG100 for halos with virial masses in the range $\log(M_{\rm vir} / \msun) = 12 \pm 0.25$. Solid lines show the TNG100 results from the 2Mpc cut and dashed lines show the halo cut. The 16-84 percentiles of the 2Mpc cut profiles are highlighted with shaded regions. Observations of the MW are shown as arrows pointing inward of errorbars placed at $\sim 100$ kpc, whereas external observations are shown as error bars at the relevant impact parameters. In the bottom panel we show the range covered by the $N_{\rm O~VIII}$ profiles published by \citet{Locatelli2024_OVIII_MW} as a brown band. Both panels show the \citetalias{Faerman2020} results as black solid lines.}
        \label{fig:o7_o8_bservations}
\end{figure}

We now compare our results for O VII and O VIII with observations, and present them in Fig. \ref{fig:o7_o8_bservations}. The top panel shows O VII and the bottom shows O VIII. As most observations focus on MW-mass galaxies, we only show TNG100 results for the appropriate mass bin ($\log (M_{\rm vir} / \msun) = 12 \pm 0.25$). The TNG100 results for O VII are shown as orange lines, and for O VIII as green lines (solid for the 2Mpc cut and dashed for the halo cut). As in Fig. \ref{fig:ovi_observations}, we indicate the 16-84 percentiles of the 2Mpc profiles using shaded regions. We also plot the O VII and O VIII profiles calculated by \cite{Faerman2020} as black solid lines.

We start with observations of O VII in the MW itself, published by \cite{Bregman2007_OVII_MW}, \cite{Gupta2012_OVII_MW}, \cite{Miller2013_MW_OVII}, \cite{Fang2015_OVII_obs}, and \cite{Das2019_OVIII_MW}. As these observations are performed ``outwards'', it is impossible to obtain column densities for given impact parameters that we can immediately compare to our TNG100 results. Alternatively, we either adopt a median column density for all observed targets as published by the authors or aggregate ourselves the given column densities (or equivalent widths that we convert to O VII column densities assuming thin lines, for which we only consider detections with SNR > 2). All observers associate their results to the inner $\sim 100$ kpc of the halo, leading us to represent each MW observation using an errorbar placed at the associated impact parameter ($\sim 0.3 R_{\rm vir}$) and an arrow pointing inwards from it. More recently, O VII columns have been measured in the CGM of external MW-like galaxies. Specifically, we include the \cite{Mathur2023_OVII_abs} Chandra observations of O VII absorption in a quasar sightline at a distance of $116 h^{-1}$ kpc from a MW-mass galaxy, and the \cite{Nicastro2023_OVII_abs} stack of both XMM-Newton and Chandra spectra of three quasars, from which O VII column densities were extracted at both UV and X-ray. 

Our TNG100 O VII profiles are in general agreement with most observations except for those of \cite{Nicastro2023_OVII_abs}, who find much higher O VII columns. The \cite{Mathur2023_OVII_abs} observation is in good agreement with the \citetalias{Faerman2020} profile, and within $\sim 0.5$ dex of the TNG100 profiles. The TNG100 profiles fall very rapidly at the inner region of the CGM, from $N_{\rm O~VII} \approx 10^{16.4} ~ \rm{cm^{-2}}$ at $b = 0.05$ down to $N_{\rm O~VII} \approx 10^{15.5} ~ \rm{cm^{-2}}$ at $b = 0.4 R_{\rm vir}$. It therefore intersects all observed MW O VII column densities, even though the broad range of impact parameters associated with them allows a lot of flexibility in comparing both results. 

As for O VIII, there are only three observations to consider, all of which are from within the MW. First is by \cite{Gupta2012_OVII_MW}, whose mean columns were calculated based on reported equivalent widths assuming thin lines. We plot the median $N_{\rm O~VIII}$. Second is by \cite{Das2019_OVIII_MW}, for which we adopt the published median $N_{\rm O~VIII}$. The third is by \cite{Locatelli2024_OVIII_MW}, who analyzed O VIII \textit{emission} lines detected in the first eROSITA All-Sky Survey. The lines were then fitted to several three dimensional models for the density distribution of hot gas in the MW, allowing them to calculate $N_{\rm O~VIII}$ as a function of impact parameter for the MW as would have been seen by an external observer. We highlight the range of O VIII column densities as a function of impact parameter allowed by all different \cite{Locatelli2024_OVIII_MW} models with a beige band. Our median TNG100 O VIII columns agrees with MW observations only for $b \lesssim 0.1 R_{\rm vir}$. At larger impact parameters the median TNG100 columns fall below the observations, but the shaded region around the TNG100 median still includes the \cite{Locatelli2024_OVIII_MW} results out to the virial radius. The shaded region surrounding the TNG100 radial profiles also envelops the \citetalias{Faerman2020} model, which in turn is in agreement with the \cite{Das2019_OVIII_MW} observation out to $\sim 0.3 R_{\rm vir}$, and with the \cite{Locatelli2024_OVIII_MW} profile out to $\sim 0.7 R_{\rm vir}$.

In general, we find TNG100 to be less consistent with O VII and O VIII observations compared to O VI observations --- except for the innermost regions of the CGM, where there is a good match.

\subsection{Half mass radii}

\begin{figure*}
        \makebox[\textwidth][c]{\includegraphics[width= \textwidth] {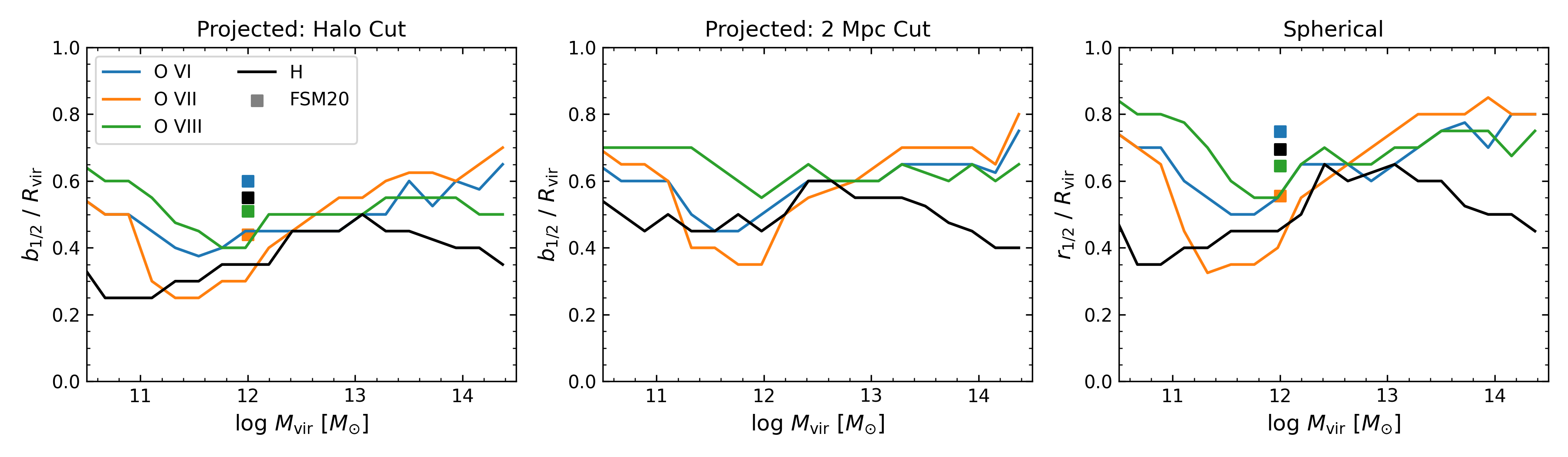}}
        \caption{The projected half mass impact parameters in units of the virial radius ($b_{1/2} / R_{\rm vir}$) estimated from the halo cut maps (left panel) and the 2Mpc cut maps (central panel), and the spherical half mass radii in units of the virial radius ($r_{1/2} / R_{\rm vir}$, right panel). The normalized $b_{1/2}$ and $r_{1/2}$ are presented for H (black lines), O VI (blue lines), O VII (orange lines), and O VIII (green lines). \citetalias{Faerman2020} half mass radii and half mass impact parameters are shown in the relevant panels, marked by black squares filled with the appropriate colors.}
        \label{fig:r_half}
    \end{figure*}

Lastly, we consider the physical distributions of O VI, O VII, and O VIII in terms of half-mass radii. We start with projected half mass impact parameters, defined for both our halo cut and for our 2Mpc cut. In both, we integrate over our maps to generate the oxygen ion masses within cylinders with impact parameter $b$:
\begin{equation} \label{eq:M_ion_proj}
    M_{\rm ion}(<b) = m_p A_{\rm O} \int_0^b{2\pi N_{\rm ion} (b') b' db'} \ \ \ .
\end{equation}
We define the half mass impact parameter, $b_{1/2}$ such that $M_{\rm ion}(<b_{1/2}) = \frac{1}{2} M_{\rm ion}(<R_{\rm vir})$. 

To calculate the spherical half mass radii, we use the particle ion number densities calculated using Eq. \ref{eq:n_ion}, and integrate over them to obtain:
\begin{equation} \label{eq:M_ion_spher}
    M_{\rm ion}(<r) = m_p A_{\rm O} \int_0^r{4\pi ~n_{\rm ion} (r') r'^2 dr'}~~~.
\end{equation}
The half mass radius, $r_{1/2}$, is defined such that $M_{\rm ion}(<r_{1/2}) = \frac{1}{2} M_{\rm ion}(<R_{\rm vir})$. In Fig. \ref{fig:r_half} we plot the normalized half mass parameters, $b_{1/2}/R_{\rm vir}$ and $r_{1/2}/R_{\rm vir}$ as functions of $M_{\rm vir}$. The left and middle panels show the projected impact parameters from the halo and 2~Mpc cuts. The right panel shows the spherical half mass radii. In these plots, O VI is in blue, O VII in orange, and O VIII in green. The black curves are for hydrogen, ionized plus neutral, as the tracer of the total CGM gas masses. We also calculated the half mass impact parameters and half mass radii for the profiles presented by \citetalias{Faerman2020}. We show these as squares at the appropriate ion identification color, and plotted in the left and right panels. The \citetalias{Faerman2020} profiles extend to only $1.1 R_{\rm vir}$, so we cannot compare their results to our 2Mpc cut.

For the oxygen ions in TNG100, $b_{1/2}/R_{\rm vir} \sim 0.5 - 0.6 $, for the halo cut maps at the lowest mass bin. The normalized impact parameters decline slightly with halo mass --- reaching minima at $M_{\rm vir} \approx 10^{11.5} ~\msun$ --- and then rise back again to $\sim 0.6 - 0.7$ for the highest mass bin. The largest variation is for O VII, for which $b_{1/2}/R_{\rm vir}$ drops to $\sim 0.3$ at $M_{\rm vir} \approx 10^{11.5} ~\msun$, and then rises to $\sim 0.7$, suggesting that for $M_{\rm vir} \approx 10^{11.5-12} ~\msun$, the O VII is localized close to the halo center. This behavior for the O VII is also present for the 2Mpc cut maps. 
In contrast, for hydrogen, $b_{1/2} / R_{\rm vir}\sim 0.3$ for the lowest mass bin, after which it gradually rises to $\sim 0.5$ at $M_{\rm vir} \approx 10^{13} ~\msun$ and then slowly drops from there. In all cases, the spherical half mass radii are $20\% \sim 25\%$ larger than the projected half mass impact parameters from the halo cut. This is not surprising in switching from projected to spherical geometries (see e.g. the two left panels of Fig. 3 in \citealp{Oren2024} for how the switch affects power law distributions). 
 
Similar trends are seen for $b_{1/2} / R_{\rm vir}$ for the 2Mpc cut maps but with slightly lower dynamical range. Moving from the halo cut to the 2Mpc cut raises $b_{1/2}/R_{\rm vir}$ by $\sim 0.2$ for 
all halo masses --- suggesting non negligible contributions for the oxygen ions from outside the virial radius. For hydrogen, however, the increase in $b_{1/2}$ from the halo cut to the 2Mpc cut changes from $\sim 50\%$ at the lowest mass bin to almost no change at the highest mass bin, suggesting that for low mass halos, a significant amount of gas is present outside of the virial radius. 

The \citetalias{Faerman2020} half mass impact parameters and half mass radii are relatively close to each other, with $0.44 < b_{1/2} / R_{\rm vir} < 0.6$ and $0.56 < r_{1/2} / R_{\rm vir} < 0.75$. 
Regarding the oxygen ions - the only difference between them are the ion fractions, shown in Fig. 6 of \citetalias{Faerman2020}. \fovi is monotonically increasing for $r \lesssim 0.5 R_{\rm vir}$, after which it remains roughly flat; \fovii sharply increases up to $\sim 0.1 R_{\rm vir}$ and then remains mostly flat (it slowly decreases by $\sim 30\%$ down to the virial radius); and \foviii sharply decreases below $\sim 0.1 R_{\rm vir}$, after which it also remains relatively flat (it slowly increases by a factor of $\sim 2$ up to the virial radius). In other words, all three ion fractions are roughly flat for $r \gtrsim 0.5 R_{\rm vir}$. Considering that $M_{\rm ion}\propto n_{\rm ion} R^3$, it is very sensitive to large radii --- and at large radii, the three ion fractions are roughly constant, leaving us with similar half mass radii for all three. 
Considering the individual ions, we find that O VII is the most concentrated, followed by O VIII and then O VI. This may be explained by the differences in the ion fractions at small radii --- \fovi rises outwards, and therefore is the least concentrated; while \foviii has a cusp at small radii, it also slowly rises outwards and is thus less concentrated than \fovii, which slowly decreases outwards. 

\cite{Stern2018_ovi} estimated the deprojected half mass radius of O VI in MW-mass galaxies observed by the COS-Halos survey \citep{Tumlinson2013_COS_halos} and by \cite{Johnson2015_OVI_survey} by smoothing the observed O VI column densities and performing an inverse Abel transform, and found it to be $0.59 ~R_{\rm vir}$ with a 0.12 dex dispersion. Considering the velocity differences allowed by observers between the galaxies and the O VI columns, this is equivalent to a half mass radius estimated within a $\sim 2.1$ Mpc sphere. In comparison, we find a projected $b_{1/2} = 0.51$ from the 2Mpc cut maps, within the error range of the result. If we assume that for the 2Mpc cut, switching from a projected $b_{1/2}$ to a spherical $r_{1/2}$ will yield a $20\%$ increase (similar to the halo cut), the equivalent half mass radius is $\sim 0.64$, well within the error range of the \cite{Stern2018_ovi} result. We should note that \cite{Stern2018_ovi} estimate $R_{\rm vir} = 190$ kpc for their sample, which is $\sim 80 \%$ of our definition of $R_{\rm vir}$ for a halo mass of $10^{12} ~\msun$ at the same redshift.

\section{Power Law Model and Thermal Phases} \label{sec:plm}
So far, we have analyzed the TNG100 column density profiles for the high oxygen ions, calculated assuming CI+PI, and we have concluded that they are consistent with available observations. We now review the power law model (PLM) for the CGM hot gas phase that we introduced in \cite{Oren2024}. As we describe below we use the PLM for analysis of our TNG100 results.

Our PLM is for the \textit{hot} phase, which in \cite{Oren2024} we defined as CGM gas with temperatures $T \ge 0.4 T_{\rm vir}$. This definition was based on our finding that for TNG100 halos, and for any virial mass, more than $90 \%$ of the thermal Sunyaev-Zeldovich electron pressure integrals (the Compton-$Y$ integrals) are built up in these hot components. We found that the radially averaged TNG100 hot CGM gas density and temperature profiles are very well represented by simple PLMs. In this model, the hydrogen gas density varies with radius, $r$, as
\begin{equation} \label{eq:PLMn}
    n_{\rm H} = n_{\rm H,vir} \left(\frac{r}{R_{\rm vir}} \right)^{-a_n} \ \ \ ,
\end{equation}
where $n_{\rm H, vir}$ is the hot gas hydrogen density at the virial radius, and the temperature as
\begin{equation} \label{eq:PLMT}
    T = \phi_T  T_{\rm vir} \left(\frac{r}{R_{\rm vir}} \right)^{-a_T} ,
\end{equation}
where $\phi_T$ is the ratio of the actual gas temperature at the virial radius to the halo virial temperature. The electron pressure in the PLM then varies as
\begin{equation}
P_e = P_{e,{\rm vir}}  \left(\frac{r}{R_{\rm vir}} \right)^{-a_{P,{\rm th}}} \ \ \ ,
\end{equation}
where $P_{e,{\rm vir}}$ is the electron pressure at the virial radius, and $a_{P,{\rm th}}=a_n + a_T$.

In \cite{Oren2024} we fit PLMs to the TNG100 CGM densities and temperatures for halos spanning $M_{\rm vir}=10^{11}$ to $3\times 10^{14}$~M$_\odot$, at $z=0$. The resulting power-law indices range from $a_n$ $\sim 1.0-2.0$, and $a_T$ $\sim 0.2 - 0.8$ (see Fig. 10 there). We also found that $\phi_T$ $\sim 0.8-1.2$, weakly dependent on mass, i.e.,~the hot gas temperatures at the virial radii are very close to $T_{\rm vir}$\footnote{Note that since $a_{\rm T} > 0$,  $\phi_T T_{\rm vir}$ is the lowest temperature for a given PLM, with temperature increasing radially inward.}. Finally, we found that the fraction of halo baryons in the hot phase, $f_{\rm hCGM}$, ranges from $\sim 0.4$ at low mass, decreasing to $\sim 0.2$ at $M_{\rm vir} \approx 10^{12.5} ~\msun$, and then rises again to $\sim 0.8$ towards $\approx 10^{14.5} ~\msun$.
The dip at $10^{12.5} ~\msun$ is due to the onset of kinetic AGN feedback in TNG \citep[see also][]{Oren2026}.

Following \cite{Oren2024}, in addition to the hot phase we define a cool phase for which $T < 3 \times 10^{4}$ K. The cool phase is characteristic of gas heated by photoionization only (mainly hydrogen and helium).  A third intermediate phase lies between cool and hot. In \cite{Oren2024} we found that the mass fraction, $f_{\rm cCGM}$, for the cool phase drops from $\sim 20 \%$ at $\approx 10^{11} ~\msun$ down to $\sim 1 \%$ at $10^{14.5} ~\msun$. For the intermediate phase, $f_{\rm itCGM} \approx 2 \%$ for most halo masses, except near $10^{12} ~\msun$ where it peaks at $\sim 10 \%$.



\subsection{Metallicity power laws}
To apply our PLM to the oxygen ion column density distributions we include a power-law form for the mass fraction in metals in the hot phase,

\begin{equation}
Z(r) = Z_{\rm vir} \left(\frac{r}{R_{\rm vir}} \right)^{-a_Z} ~~ ,
\label{eq:PLMZ}
\end{equation}
where $Z_{\rm vir}$ is the mass fraction at the virial radius.
Integration over the volume must give the total metal mass in the hot CGM, i.e.,

\begin{equation}
M_{\rm Z,hCGM} = \mu_H \int n_H(r) Z(r) dV ~~.
\end{equation}
where $\mu_H \approx 1.37 m_p = 2.29 \times 10^{-24}$ gr is the mean mass per proton in a fully ionized medium. Therefore, our Eq.~\ref{eq:PLMZ} is for the mass weighted metallicity. In our procedure we first compute the mass weighted metallicity in radial shells through the TNG100 halos, which we then fit to a power law distribution to obtain $a_Z$. We also set an inner boundary, $r_0$, for the metallicity power-law distribution. Thus,
\begin{equation}
Z_{\rm vir} = \frac{3 - a_n - a_Z}{3 - a_n} \frac{1 - x_0^{3 - a_n}}{1 - x_0^{3 - a_n - a_Z}} \frac{M_{\rm Z, hCGM}}{M_{\rm hCGM}} ~~,
\end{equation}
where $x_0 \equiv r_0 / R_{\rm vir}$. We set $x_0=0.04$, given that the half-mass-radii of $z=0$ TNG100 galaxies are typically at $\sim 0.02 R_{\rm vir}$ \citep{Karmakar2023}. Unlike the normalization factors for the density and temperature power-laws, for our metallicity PLM we include an inner boundary to prevent $Z_{\rm vir}$ from becoming negative in the rare cases where $a_n + a_Z > 3$ (this in fact occurs for 39 of our 880 halo sample, all of which have $M_{\rm vir} < 10^{11} ~\rm{M_{\odot}}$). 

For our PLM the mean (mass-weighted) metallicity within the CGM is 
\begin{equation} \label{eq:mean_Z_hot}
\langle Z \rangle = \frac{3 - a_n}{3 - a_n - a_Z} Z_{\rm vir}~~.
\end{equation}
The different PLM parameters are functions of halo mass. For $a_n$ we use the values from \cite{Oren2024}, shown in Fig.~10 there. We now present and discuss the metallicity parameters we infer from the simulation. In our figures and discussion below we present results in terms of the normalized metallicity 
\begin{equation}\label{eq:metal_abundance}
    Z^{\prime} \equiv Z / 0.013~~,
\end{equation}
relative to the solar metal mass fraction from \citet{Asplund2009_sun}.

\begin{figure*}
        \makebox[\textwidth][c]{\includegraphics[width= \textwidth] {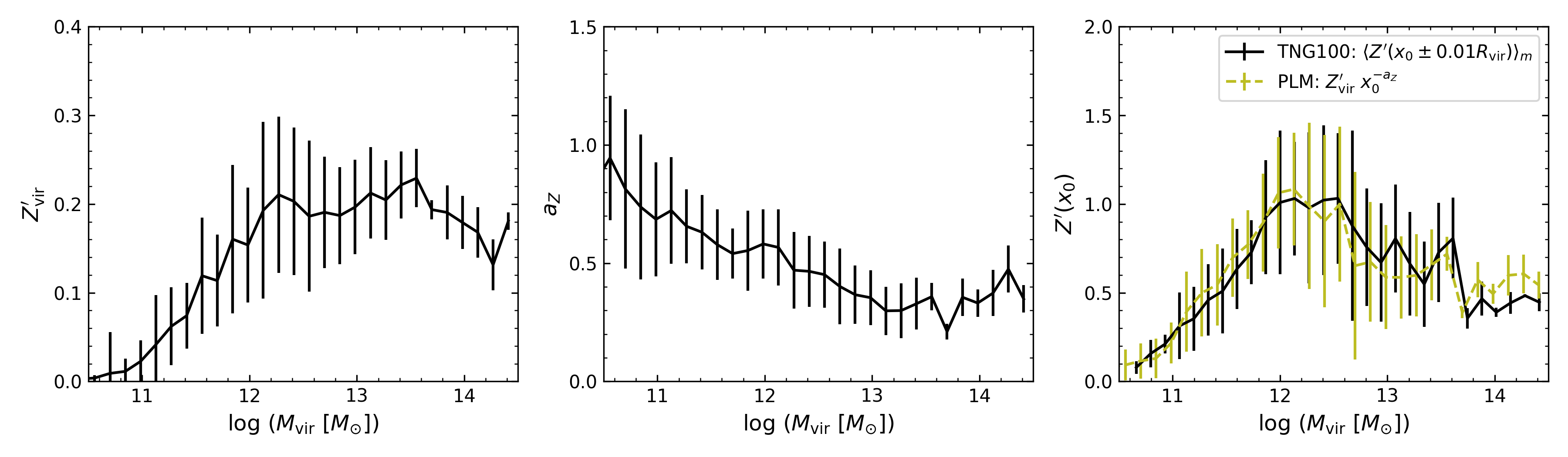}}
        \caption{Left: The metallicity, $Z^{\prime}_{\rm vir}$ (relative to Solar) at the virial radius, as a function of $M_{\rm vir}$. Middle: the PLM metallicity index, $a_{\rm Z}$ (see Eq.~\ref{eq:PLMZ}). Right: The metallicity $Z^{\prime}_{\rm vir}$ at the innermost CGM radius, $x_0 \equiv r_0 / R_{\rm vir}$. Black is the TNG100 mass weighted metallicity at $x_0$ in solar units, and yellow is the PLM prediction, showing that the metallicity of the CGM in TNG100 galaxies is well approximated by a power law of the radius.}
        \label{fig:Z_PLM}
\end{figure*}

We show our PLM fits to the TNG100 metallicities in Fig. \ref{fig:Z_PLM}.
The left panel shows that $Z_{\rm vir}^{\prime}$ first grows with halo mass and stabilizes at $Z_{\rm vir}^{\prime}\sim 0.2$ at $M_{\rm vir} \gtrsim 10^{12.3} ~\rm{M_{\odot}}$, and then decreases for $M_{\rm vir} \gtrsim 10^{13.5} ~\msun$. This behavior is similar to what was found by \cite{Oren2026} for the CGM metallicities in TNG100 halos, and attributed to the onset of kinetic AGN feedback and its relation to the gravitational potential of the halo\footnote{The onset of kinetic AGN feedback in IllustrisTNG halts star formation and the generation of fresh metals, alongside ejection of existing metals from the galaxy into the CGM. Evidence for this has been found in the stellar-mass metallicity ratio (MZR, \citealp{Torrey2019_MZR}), or in the metal flow rates through both the ISM-CGM boundary and the CGM-IGM boundary \citep{Oren2026}. The drop in larger masses is attributed to the gravitational potential of massive host halos preventing AGN-driven winds from reaching the virial radius, effectively lowering the metal fraction at this scale.}. The middle panel of Fig. \ref{fig:Z_PLM} shows that the metallicity power-law slope $a_{Z}$ decreases with halo mass. This reflects a picture in which the CGM metallicity gradually becomes more uniform as halos grow and strong feedback vigorously mixes their gas content. 

To test the accuracy of our metallicity PLM, in the right panel of Fig. \ref{fig:Z_PLM} we compare the CGM metallicity at $x_0 = 0.04 R_{\rm vir}$ from our PLM and the actual mass weighted CGM metallicity in a shell ranging from $0.03 R_{\rm vir}$ to $0.05 R_{\rm vir}$. The TNG100 results and our PLM are in good agreement, staying within $20 \%$ of each other for all halo masses, suggesting that a power law of the radius is indeed a good approximation for the metallicity of the CGM surrounding TNG100 galaxies. The TNG100 results are cut off for $\log(M_{\rm vir} / M_{\odot}) \lesssim 10.8$, because for these low masses no CGM particles are found in a $0.02 R_{\rm vir} \approx 2$ kpc shell that lies very close to the galaxy.



\subsection{Volume filling factors of the different CGM phases}
\begin{figure}
        \includegraphics[width=0.45 \textwidth]{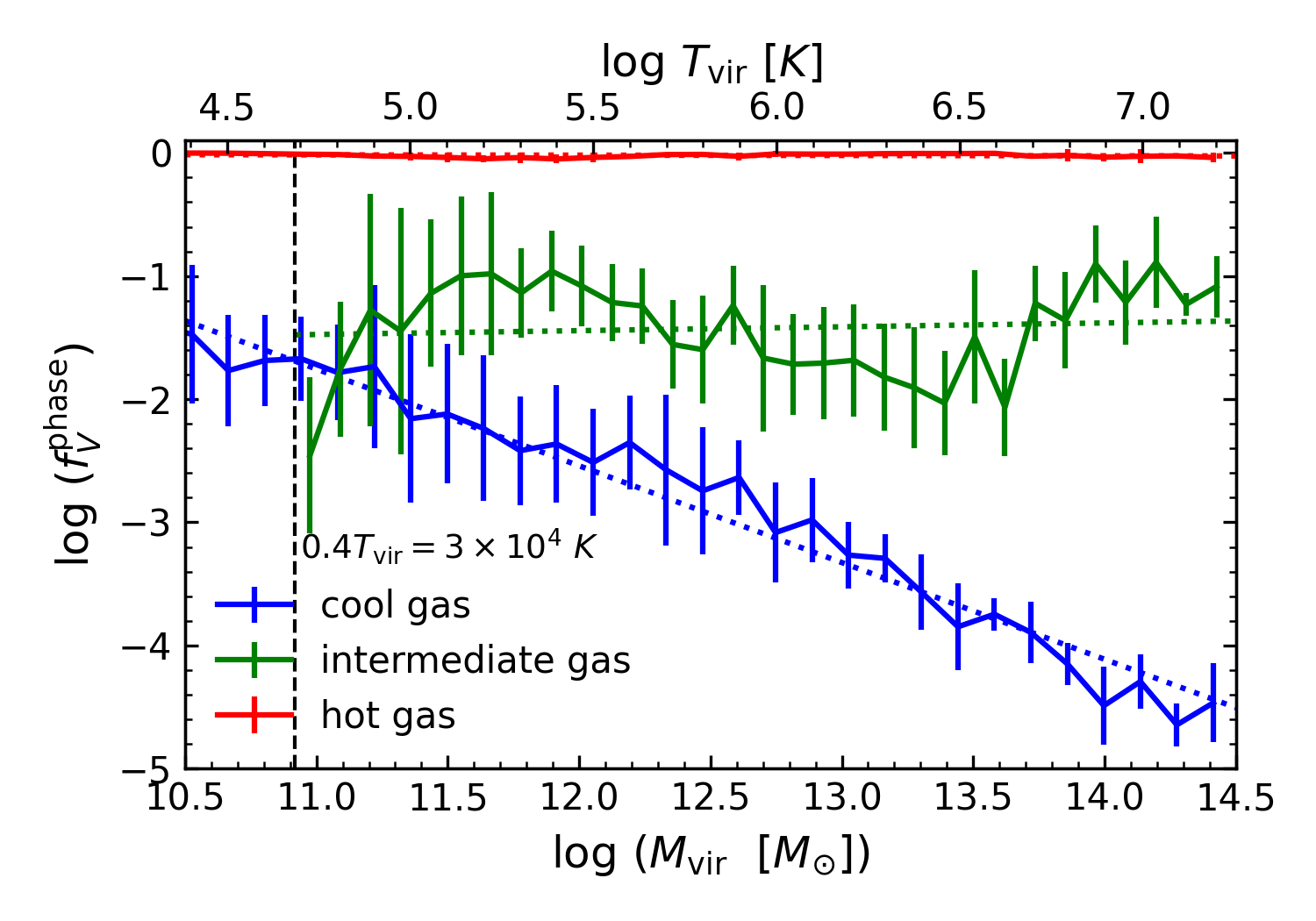}
        \caption{Volume filling factors of the cool (blue), intermediate (green), and hot phase (red) of the CGM, as a function of halo mass and virial temperature (upper $x$-axis). Error bars show the standard deviations. The dotted lines show the linear fits between $\log(f_{\rm V})$ and $\log(M_{\rm vir})$. The vertical dashed black line marks the virial mass where $0.4 T_{\rm vir} = 3 \times 10^{4}$ K, below which there is no intermediate phase.}
        \label{fig:f_V_mvir}
\end{figure}

We now consider the volume filling factors of our three gas components, hot, intermediate, and cool. We define the volume filling factor, $f_{\rm V}$, as the volume occupied by the gas component within the CGM normalized by the total CGM volume. In Fig. \ref{fig:f_V_mvir} we plot the median volume filling factors as functions of virial mass, for our sample of TNG100 galaxies. The red curves are for the hot phase, green is intermediate, and blue is for cool. We also show linear fits (dotted lines) for $\log(f_V)$ versus $\log (M_{\rm vir} / 10^{12} \msun)$. We note that for the intermediate phase no results are shown for $M_{\rm vir} \le 8.3 \times 10^{10} ~\msun$, because for this virial mass $0.4 T_{\rm vir} = 3\times 10^{4} $ K, and no gas is associated with the intermediate phase below this halo mass.

We find,
\begin{itemize}
    \item[$\bullet$] $\log(f_{\rm V}^{\rm cool}) = - 0.786 ~ \log(M_{\rm vir} / 10^{12} ~\msun) - 2.541$
    \item[$\bullet$] $\log(f_{\rm V}^{\rm it}) = 0.031 ~ \log(M_{\rm vir} / 10^{12} ~\msun) -1.444$
    \item[$\bullet$] $\log(f_{\rm V}^{\rm hot}) = -0.003 ~ \log(M_{\rm vir} / 10^{12} ~\msun) - 0.019$
\end{itemize}

For the hot gas $f^{\rm hot}_V$ is very close to unity for all halo masses. For intermediate temperature gas $f^{\rm it}_V$ is roughly constant, at $0.03 -0.04$ for $M_{\rm vir}$ from $10^{11}$ to $10^{14.5} ~\msun$. For cool gas, $f^{\rm cool }_V$ decreases dramatically from $\sim 0.03$ at the lowest mass bin down to $\sim 3\times 10^{-5}$ at the highest masses.  


It is interesting to compare these results to estimates obtained from observations. For example, \cite{Faerman2023_Mcool} present a model for the cool CGM and apply it to the HI and low metal ions column densities measured in the COS-Halos survey \citep{Werk2013, Tumlinson2013_COS_halos, Prochaska2017}. A fiducial model with $f_{\rm V}^{\rm cool} \approx 0.013$ is consistent with the average column densities, and has a cool gas mass of $M_{\rm cool} \approx 3\times 10^{9} ~\msun$ (equivalent to $f_{\rm cCGM} = 0.02$). However, the measured columns have significant scatter, and \cite{Faerman2023_Mcool} show that models with $0.005 < f_{\rm V}^{\rm cool} < 0.045$, and gas masses of $10^{9} <M_{\rm cool}/\msun < 10^{10}$ reproduce most of the observations. These estimates are lower than the mass inferred by \citet{Prochaska2017} for the same data set, of $M_{\rm cool} =  (9.2 \pm 4.3) \times 10^{10}~\msun$. In comparison, MW-mass galaxies in TNG100 have $f^{\rm cool}_{\rm V} \approx 3\times 10^{-3}$ and cold gas masses of $M_{\rm cCGM} \approx 1.6\times 10^{10} ~ \msun$ (or $f_{\rm cCGM} \approx 0.1$), between the estimates by P17 and FW23\footnote{Compared to FW23, the TNG100 cold gas masses are higher while their volume filling factors are lower. This is a result of the gas densities in the simulation being higher than in the FW23 model, which allows lower gas densities motivated by non-thermal pressure support or (thermal) pressure imbalance between the cool and warm/hot ambient phases.}.

More recently, \cite{Faerman2025_Mcool} model the HI column densities in the CGM of dwarf galaxies, measured and compiled by \cite{Zheng2024} and \cite{Mishra2024}. They explore two different scenarios for the cool gas: clumpy, for which $f_{\rm V}^{\rm cool} = 0.01$, or volume-filling, for which $f_{\rm V}^{\rm cool} = 1$. For a halo mass of $M_{\rm vir} = 10^{11} ~\msun$ they find $M_{\rm cCGM} \approx 10^{8} ~\msun$ for the clumpy model, or $M_{\rm cCGM} \approx 10^{9} ~\msun$ for the volume-filling model\footnote{As shown in \cite{Faerman2025_Mcool}, this scenario provides an upper limit for the cool gas mass that reproduces the measured HI column densities.} (corresponding to $f_{\rm cCGM} \approx 0.006$ or $f_{\rm cCGM} \approx 0.06$, respectively). In comparison, the volume filling fraction of the cool gas in TNG100 halos with $M_{\rm vir} = 10^{11} ~\msun$ is $f_{\rm V}^{\rm cool} \approx 0.02$, but their median cool gas mass is $M_{\rm cCGM} = 2.5 \times 10^{9} ~\msun$ (or $f_{\rm cCGM} = 0.16$). So while the volume filling factor of TNG100 cold gas is very close to the clumpy model used in \cite{Faerman2025_Mcool}, in practice the mass of the cool gas in the CGM of TNG100 $10^{11} ~\msun$ halos is even higher than the cold gas mass obtained by the model that assumes the cold gas is volume filling.

\section{Mean Oxygen Column densities} \label{sec:mean_columns}
The mean column density of an ion within the CGM,
\begin{equation} \label{eq:mean_N}
    \langle N_{\rm ion} \rangle = {\cal N}_{\rm ion} / \pi R_{\rm vir}^2 \ \ \ ,
\end{equation}
is a useful parameter with which to estimate expected absorption line strengths, as well as to determine the primary ionization mechanisms (collisional ionization versus photoionization) and the phases, hot, intermediate, or cool, within which the ion is produced. In Eq.~(\ref{eq:mean_N}), ${\cal N}_{\rm ion}\equiv M_{\rm ion} / m_p A_{\rm O}$ is the total number of ion particles (O VI, O VII, or O VIII) within the virial radius, and $\pi R_{\rm vir}^2$ is the CGM cross section. The mean columns provide a single number per halo (per ion), which can be computed and easily presented for large samples of simulated halos, and the results provide compact information on the predicted ion abundances versus halo mass.

We compute $\langle N_{\rm ion} \rangle$ in several ways. First, directly using our TNG100 post-processing results, assuming either CI+PI, or only CIE. In these computations we include the contributions for a given ion from all phases, hot, intermediate, and cool. Second, we also compute the ionization states using our PLM representations for the density, temperature, and metallicity structures of the hot components. We then supplement the PLMs with an analytic approximation for the mean columns produced in the intermediate temperature phase. Comparing the results of these four methods enables us to determine the dominating ionization mechanisms and the thermal phases within which a given ion is produced.

\subsection[Mean ion column densities in TNG100]{\Nion in TNG100} \label{subsec:NionTNG100}
To calculate the mean ion column densities using our TNG100 results we use the elemental oxygen mass fractions output by IllustrisTNG and the Cloudy-computed ion fractions in order to obtain the total ion mass 
\begin{equation} \label{eq:M_ion}
M_{\rm ion} = \sum_{\rm CGM} m_{\rm ion,i} = \sum_{\rm CGM} m_{\rm gas,i} \left(\frac{m_{\rm O}}{m_{\rm gas}}\right)_i f_{\rm ion,i} ~~~,
\end{equation}
where the sum is over all CGM particles within the virial radius (cool, intermediate and hot), and $m_{\rm gas,i}$ is the gas mass of an individual CGM particle in TNG100. 
 
We compute the ion fractions, $f_{\rm ion,i}$ in two ways. First assuming pure collisional ionization equilibrium (see Fig. \ref{fig:f_ion_CIE}), and second via collisions plus photoionziation by the metagalactic field (see Fig. \ref{fig:f_ion_full}). As we have stated, we refer to these options as CIE and CI+PI. If the results for CI+PI are close to CIE this implies that photoionization is a minor effect or negligible. If CI+PI yields mean columns larger than for CIE this implies photoionization is significant or even dominant. A third possibility is for CI+PI to reduce the mean columns below CIE. This implies photo-destruction of collisionally ionized gas. However, as we discuss below, this is generally a small effect within the CGM of the TNG100 galaxies. We summarize the three possibilities in Table~\ref{table_ionization}. If the mass ratio $M_{\rm CIE}/M_{\rm CI+PI} \le 0.5$ photoionization dominates, if this mass ratio is within 0.5 and 2, we say that collisional ionization dominates, and $M_{\rm CIE}/M_{\rm CI+PI} \ge 2$ is a diagnostic of photodestruction. 

\begin{table} 
\centering
\caption{Our definitions for the dominant ionization mechanism of an ion species within the CGM.}
\label{table_ionization}
\begin{tabular}{@{}l|l@{}}
\toprule
\textbf{Case}          & \textbf{$M_{\rm CIE}$ to $M_{\rm CI+PI}$ ratio} \\ \midrule
Photoionization      & $M_{\rm CIE} / M_{\rm CI+PI} \le 0.5$           \\
Collisional ionization & $0.5 < M_{\rm CIE} / M_{\rm CI+PI} \le 2$       \\
Photodestruction        & $2 < M_{\rm CIE} / M_{\rm CI+PI}$               \\ \bottomrule
\end{tabular}
\end{table}

\begin{figure*}
        \makebox[\textwidth][c]{\includegraphics[width= \textwidth]{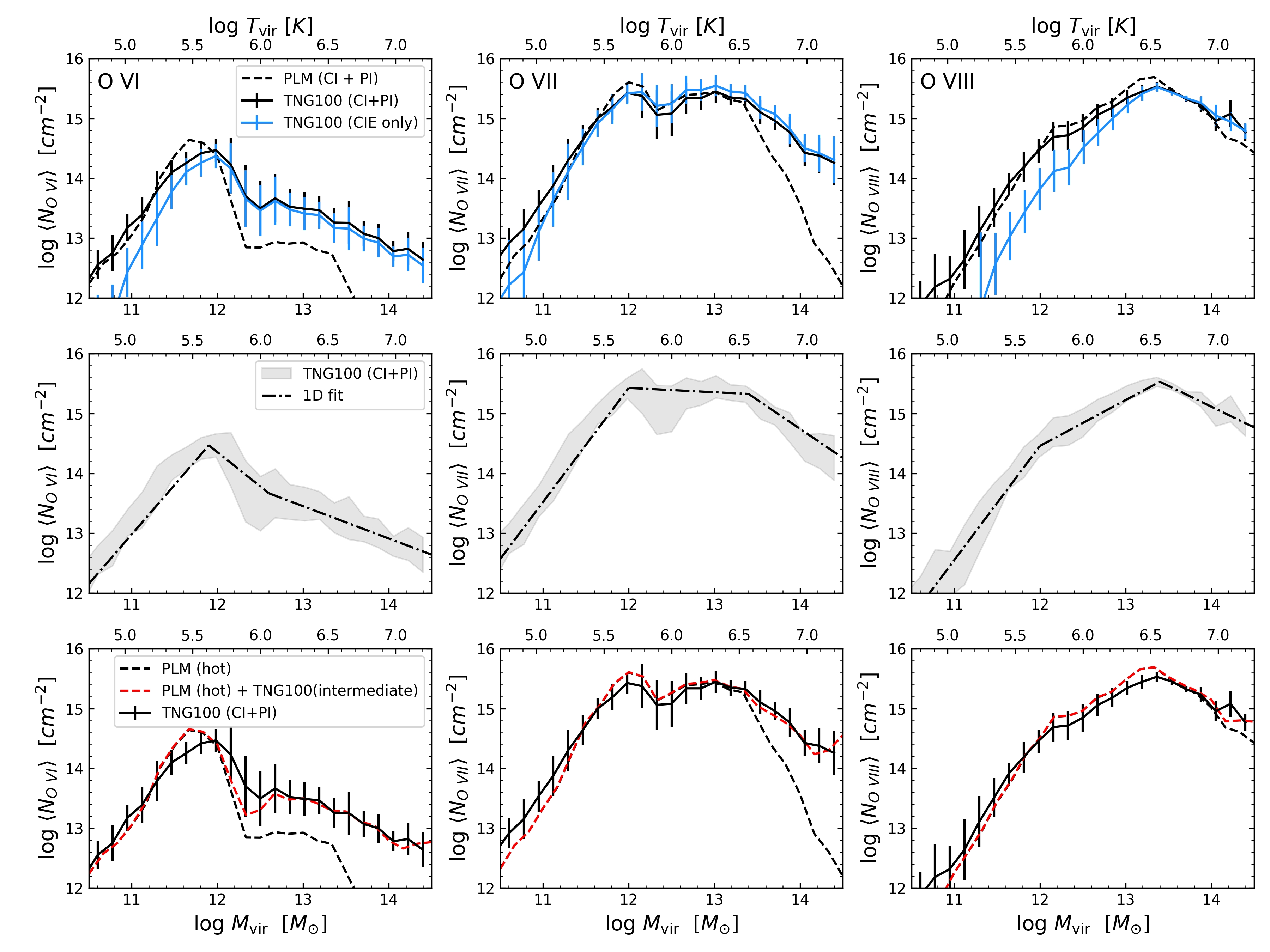}}
        \caption{Mean column densities, O VI (left), O VII (middle) and O VIII (right), as functions of halo mass and virial temperature (upper $x$-axis). 
        Top row: Solid curves are for our TNG100 halo sample, assuming collisional plus photoionization (black) or just collisional ionization equilibrium (blue). The black dashed curves are the hot phase PLM results (see text). 
        Middle row: Shaded region is for the 16-84 percentile of the TNG100 (CI+PI) mean column densities, and the dashdot line is the piece-wise linear fit to the median.
        Bottom row: In addition to the TNG100 (CI+PI) black error-bars and the hot phase PLM prediction black dashed lines, the red dashed lines show the hot gas PLM prediction with the correction for the intermediate phase (see text in \S~\ref{sec:it_contribution}). }
        \label{fig:plm_v_tng}
    \end{figure*}

In Figure \ref{fig:plm_v_tng} we show the mean columns for our three oxygen ions, as functions of $M_{\rm vir}$. We have carried out the calculations for 25 mass bins. In the upper row of Fig.~\ref{fig:plm_v_tng} the solid black curves are for CI+PI, and blue are for CIE only. (The dashed black curves are for the PLM computations we describe below). Our results for the mean columns as functions of halo mass are similar to and consistent with the computations presented by \cite{Nelson2018_TNG_absorption} (see their Fig.8).

To a first approximation the shapes of the curves in Fig.~\ref{fig:plm_v_tng} resemble the pure CIE ion fraction curves in Fig.~\ref{fig:f_ion_CIE}. For example, $\langle N_{\rm O~VI}\rangle$ peaks for halos with $T_{\rm vir}\approx 3\times 10^5$~K, corresponding to temperatures at which the O VI CIE ion fractions are largest \citep[see also][]{Oppenheimer2016_ovi_EAGLE}. The peak O VI columns are $\sim 3\times 10^{14}$~cm$^{-2}$, comparable to the observed column densities discussed above, and again indicating that O VI is produced naturally for halos with masses near $10^{12}$~M$_\odot$. Similarly, \Novii is largest for $T_{\rm vir}$ between $3\times 10^5$ and $3\times 10^6$~K, as expected for CIE, with maximal mean columns near $3\times 10^{15}$~cm$^{-2}$. Finally, \Noviii peaks at $3\times 10^{15}$~cm$^{-2}$ at $T_{\rm vir} \sim 3\times 10^{6}$ K, as expected for CIE as seen in Fig.~\ref{fig:f_ion_CIE}. 

However, the differences between the (black) CI+PI and (blue) pure CIE curves, especially at masses to the left of the CIE peaks, show that photoionization becomes increasingly important with decreasing virial mass. This is most noticeable for O VIII, for which the (blue) CIE curve falls well below the (black) CI+PI curve for $M_{\rm vir} < 10^{13.4}$~M$_\odot$. For example, at $M_{\rm vir}=10^{12}$~M$_\odot$ it is evident that O VIII is produced mainly by photoionization. At such masses the gas is not hot enough to produce O VIII collisionally. For masses to the right of the CIE peaks the CI+PI and CIE curves overlap, indicating that collisional ionization dominates in these regimes.

\begin{figure}
        \includegraphics[width=0.45 \textwidth]{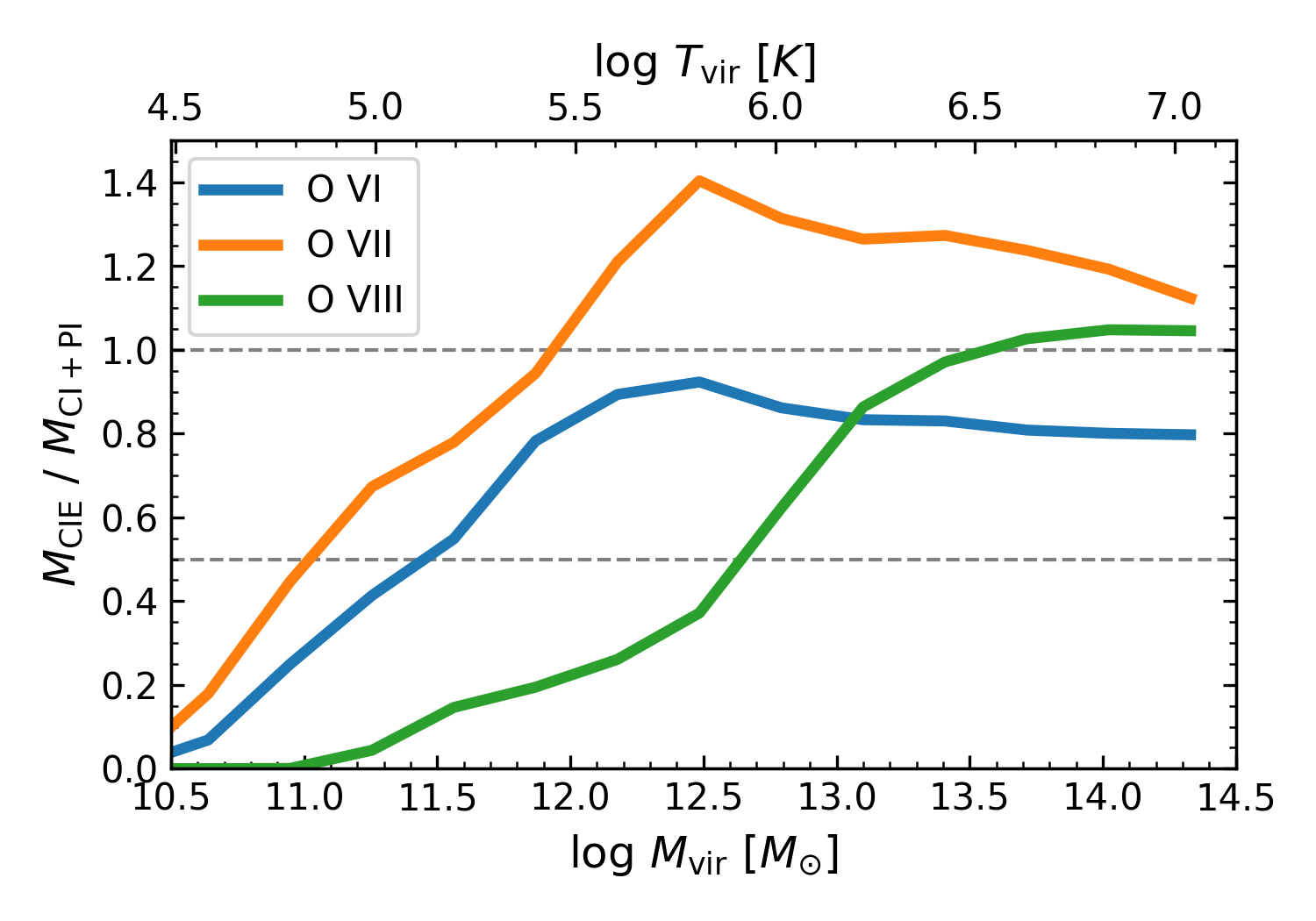}
        \caption{The ratio, $M_{\rm CIE}/M_{\rm CI+PI}$, of the CGM ion masses calculated assuming CIE to the ion masses calculated assuming CI+PI. Orange is O VI, blue is O VII, and green is O VIII. }
        \label{fig:CIE_fraction}
\end{figure}

For additional quantification of the relative roles of photoionization versus collisional ionization, 
in Fig.~\ref{fig:CIE_fraction} we plot the ratios, $M_{\rm CIE}/M_{\rm CI+PI}$ of the ion masses, assuming CIE in the numerator and CI+PI in the denominator, as functions of halo virial mass. At low $M_{\rm vir}$ the CIE to CI+PI ratios are much less than unity for all three ions, indicating that photoionization dominates the ion formation at sufficiently low virial mass. Fig.~\ref{fig:CIE_fraction} shows that $M_{\rm CIE}/M_{\rm CI+PI} > 0.5$ at virial masses of $10^{11.5}$, $10^{11}$, and $10^{12.5}$~M$_\odot$, for O VI, O VII, and O VIII respectively. These are the virial masses at which collisional ionization begins to dominate the production of each ion. The CIE to PI+CI ratio for O VI (blue curve) decreases slightly below unity at large $M_{\rm vir}$, indicating that photoionization still contributes somewhat to the O VI production at these masses. For O VII, the CIE to PI+CI ratio (orange curve) rises above unity implying that photodestruction plays a non-negligible role in removing the O VII at high virial masses. However, photodestruction remains subdominant compared to collisional ionization as a removal mechanism because $M_{\rm CIE}/M_{\rm CI+PI} < 2$ everywhere.

For convenient representations of the mean oxygen ion columns as functions of halo virial mass we have fit our CI+PI results as three-piece functional forms, given by
\begin{equation}
log(\langle N_{\rm ion}\rangle / \rm{cm^{-2}}) = \alpha \times log(M_{\rm vir} / 10^{12} ~\msun) + \beta \ \ \ ,
\end{equation}
The mass ranges and fit parameters $\alpha$ and $\beta$ are listed in Table \ref{table_N_M_fit}.
We plot these fitting functions in the middle row of Fig. \ref{fig:plm_v_tng}.

\begin{table} 
\centering
\caption{Results of the linear fitting procedure of the TNG100 mean column densities assuming CI+PI as a function of halo virial mass, divided into three mass segments and of the form 
\newline $\log(\langle N_{\rm ion} \rangle / cm^{-2}) = \alpha \times \log(M_{\rm vir} / 10^{12} M_{\odot}) + \beta$.} \label{table_N_M_fit}
\begin{tabular}{c|cc}
\toprule

$\log(M_{\rm vir} / M_{\odot})$ range & $\alpha$ & $\beta$ \\
\midrule

\multicolumn{3}{c}{\textbf{O VI}} \\
\midrule
10.5 --- 11.9 & 1.65 & 14.63 \\
11.9 --- 12.6 & -1.14 & 14.35 \\
12.6 --- 14.5 & -0.54 & 13.99 \\

\midrule
\multicolumn{3}{c}{\textbf{O VII}} \\
\midrule
10.5 --- 12.0 & 1.91 & 15.43 \\
12.0 --- 13.4 & -0.07 & 15.43 \\
13.4 --- 14.5 & -0.97 & 16.69 \\

\midrule
\multicolumn{3}{c}{\textbf{O VIII}} \\
\midrule
10.5 --- 12.0 & 1.91 & 14.46 \\
12.0 --- 13.4 & 0.77 & 14.46 \\
13.4 --- 14.5 & -0.69 & 16.50 \\

\bottomrule
\end{tabular}
\end{table}

\subsection[Estimating mean ion column densities using a PLM]{\Nion with the PLM} \label{subsec:mean_N_PLM}
We now compute the mean column densities using our PLM for the hot phase. Comparison to the direct TNG100 results will show whether and when the intermediate temperature and cool gas components contribute to the ion masses and mean columns.

To calculate the mean column density using our PLM, we first estimate the ion density (cm$^{-3}$) at a given radius using:
\begin{equation} \label{eq:n_ion_plm}
    n_{\rm ion}(r) = n_H(r) a_{\odot} Z^{\prime}(r) f_{\rm ion}[n_H(r), T(r)] \ \ \ ,
\end{equation}
where $n_H(r)$, $T(r)$, and $Z^{\prime}(r)$ are given by Eqs.~(\ref{eq:PLMn}), (\ref{eq:PLMT}) and (\ref{eq:PLMZ}), and $a_{\odot} = 4.9\times10^{-4}$ is the Solar oxygen abundance by number \citep{Asplund2009_sun}. Here, $f_{\rm ion}$ is the ion fraction assuming CI+PI given the PLM values for $n_H$ and $T$. We compute the median PLM parameters $a_{P,{\rm th}}$, $a_n$, $a_Z$, $f_{\rm hCGM}$, $\phi_T$, and $Z^{\prime}_{\rm vir}$, for each of the 25 mass bins and for use in Eq.~(\ref{eq:n_ion_plm}).

We then calculate the total number of ion particles within the virial radius using:
\begin{equation}\label{eq:N_ion_plm}
    {\cal N}_{\rm ion}^{\rm hot} = 4 \pi \int_{r_0}^{R{\rm vir}} n_{\rm ion}(r) r^2 dr,
\end{equation}

enabling us to estimate the mean column density using 
Eq.~(\ref{eq:mean_N}). The superscript ``hot'' indicates that this is the particle number for the hot phase as represented by the PLM. We only consider CI+PI, and the results are the black dashed curves in 
Fig.~\ref{fig:plm_v_tng}.

For low halo masses, to the left of the CIE peaks, the PLMs provide good estimates for the TNG100 mean columns for all three ions. However, at high halo masses the PLM columns fall far below the TNG100 predictions. This is because at high masses the ions are not produced in the hot ($T > 0.4T_{\rm vir}$) virialized gas as represented by the PLMs, but rather in cooler gas near the CIE peak temperatures, which is actually in the intermediate regime for these halo masses.  

\begin{figure*}
        \makebox[\textwidth][c]{\includegraphics[width= \textwidth]{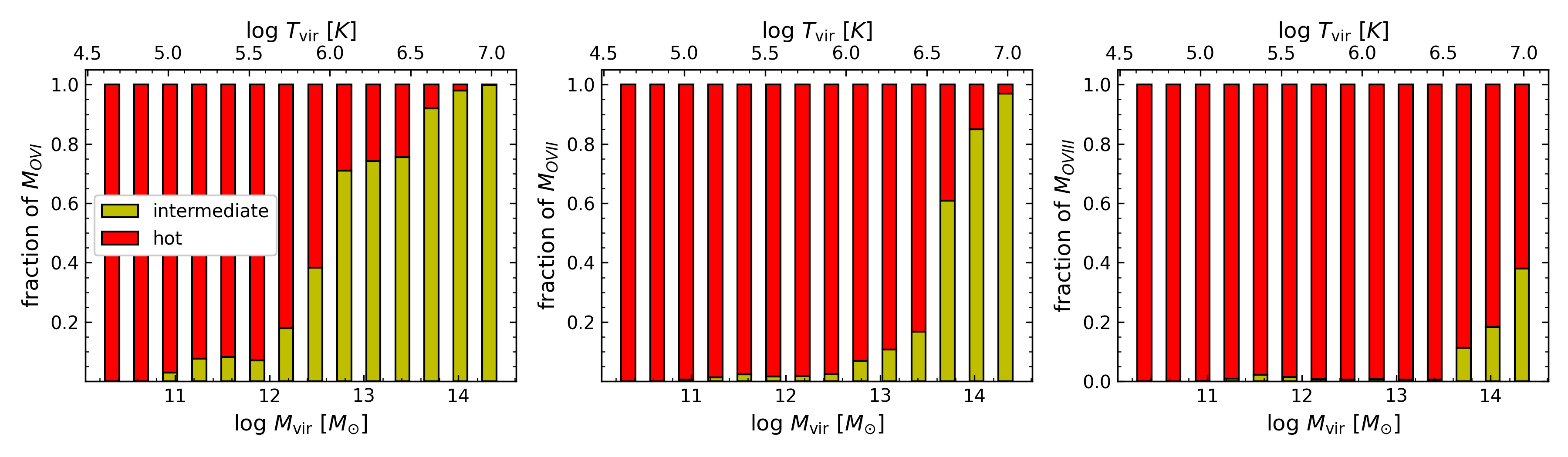}}
        \caption{Oxygen ion mass fractions in the hot (red) and intermediate temperature (yellow) gas phases, as functions of halo mass for our TNG100 sample, assuming CI+PI (see text).}
        \label{fig:mass_fractions}
    \end{figure*}

In Fig. \ref{fig:mass_fractions} we show the mass fractions produced in the hot (red) versus intermediate temperature (yellow) phases, for the three ions (left, middle, and right panels). At low halo masses with low virial temperatures, the ions are generated formally within the hot phases for these halos, even though photoionization dominates the ion production. At larger halo masses, and as the virial temperatures exceed the peak CIE temperatures, increasing mass fractions are produced collisionally in the intermediate phases which now bracket gas near the CIE peak temperatures. The ion masses associated with the intermediate phase rise above 10\% of the total ion mass within the virial radius at $\log(M_{\rm vir} / M_{\odot}) \approx 12$ for O VI, $\log(M_{\rm vir} / M_{\odot}) \approx 13.4$ for O VII, and $\log(M_{\rm vir} / M_{\odot}) \approx 14$ for O VIII. These are the masses where the PLM curves start falling below the TNG100 curves for \Nion in Figure \ref{fig:plm_v_tng}. For these ions the mass fractions in the cool phase are negligible, and not shown in Fig. \ref{fig:mass_fractions}.

\subsection{Intermediate phase} \label{sec:it_contribution}
So far, we have shown that at sufficiently low halo masses the ions are produced by photoionization, with growing contributions by collisional ionization with increasing halo mass. The PLM captures this behavior so long as the ions are mainly produced in the hot phase for a given $M_{\rm vir}$ and $T_{\rm vir}$, independent of the ionization mechanism. However as is seen in Fig.~\ref{fig:plm_v_tng} at high masses the PLM predictions fall well below the full TNG100 results, especially for O VI. The reason is that at high masses the collisionally ionized ions are produced in the intermediate phases, which bracket the CIE peak temperatures, rather than in the hot gas represented by the PLM. The hot gas is too hot.

We can account for the contributions from the intermediate phase in two ways, as follows. First, we directly compute the ion masses, and ion particle number in the intermediate phase, ${\cal N}^{\rm it}_{\rm ion}$, using Eq.~\ref{eq:M_ion} but summing only over gas particles in the intermediate temperature range. In doing this we assume CI+PI for the ion fractions $f_{\rm ion}^{\rm it}$. 

Second, we make the approximation
\begin{equation}
    {\cal N}^{\rm it}_{\rm ion} = \frac{M_{\rm itCGM}}{\mu_H} a_\odot \langle Z^{\prime}\rangle^{\rm it} \langle f_{\rm ion,CIE}\rangle ^{\rm it} ~~,
    \label{eq:Nionit}
\end{equation}
where $M_{\rm itCGM}$ is the CGM gas mass in the intermediate phase, $a_\odot=4.9\times 10^{-4}$ is the Solar abundance of oxygen, $\langle Z^{\prime}\rangle^{\rm it}$ is the mass weighted mean metallicity of the intermediate phase, and $\langle f_{\rm ion}\rangle ^{\rm it}$ is the mean CIE ion fraction associated with the intermediate phase.
The mass $M_{\rm itCGM} = M_{\rm vir} f_b f_{\rm itCGM}$, where $f_b=0.156$ is the cosmic baryon fraction, and $f_{\rm itCGM}$ is the fraction of available baryons in the intermediate phase. In evaluating Eq.~\ref{eq:Nionit} we set $f_{\rm itCGM}=0.02$ for all halo masses, as appropriate for $M_{\rm vir} \gtrsim 10^{12.5}$~M$_\odot$ \cite[see Fig.8 of][]{Oren2024}. 

\begin{figure*}
        \makebox[\textwidth][c]{\includegraphics[width= \textwidth]{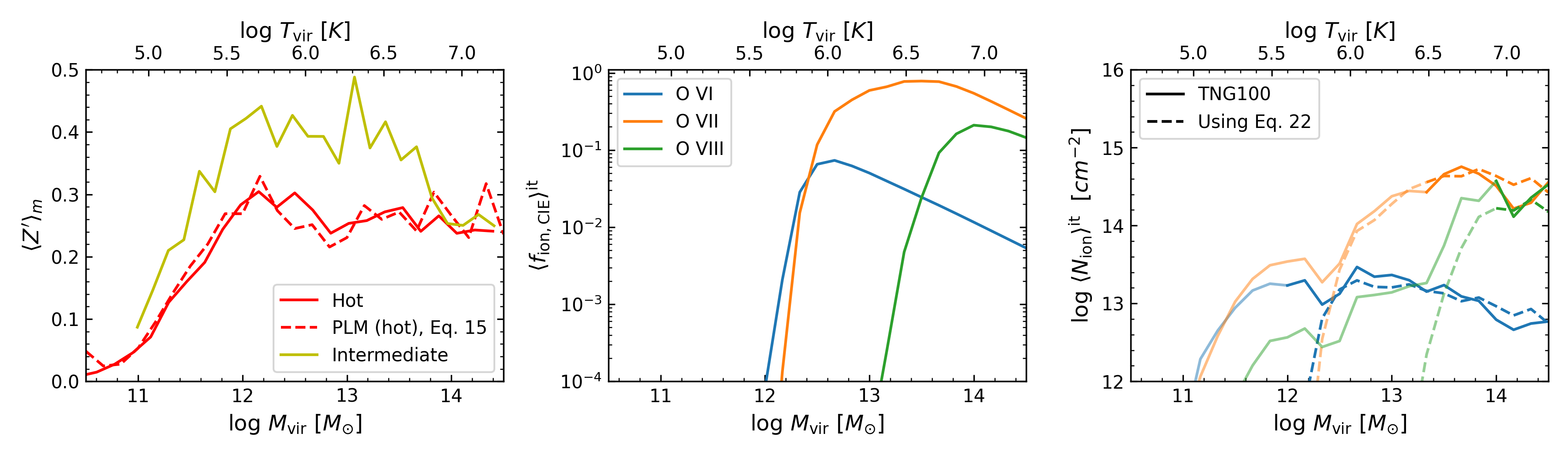}}
        \caption{Left: Mass weighted metallicity in the intermediate phase of the CGM in TNG100 (yellow) compared to the mass weighted metallicity in the hot phase of the CGM in TNG100 (solid red) and the the hot phase mass weighted metallicity obtained by our PLM (dashed red).
        Center: Oxygen ion fractions in the intermediate temperature range, $\langle f_{\rm ion,CIE}\rangle^{\rm it}$, assuming only CIE, as a function of temperature and its associated virial mass.
        Right: The mean column densities associated with the intermediate phase alone, $\langle N_{\rm ion} \rangle^{\rm it}$, from TNG100 (solid lines) and our approximated approach (dashed lines). The results are highlighted in halo masses where the contribution from the intermediate phase rises above $10\%$ (see Fig. \ref{fig:mass_fractions}).}
        \label{fig:IT_components}
    \end{figure*}

In Fig. \ref{fig:IT_components} (left panel) we plot the mass weighted metallicities $\langle Z^\prime \rangle^{\rm hot}$ and $\langle Z^\prime \rangle^{\rm it}$ for the hot and intermediate components given the TNG100 CGM particle data.  The dashed curve in Fig. \ref{fig:IT_components} is $\langle Z^\prime \rangle^{\rm hot}$ using the PLM expression Eq.~\ref{eq:mean_Z_hot} together with the PLM parameters for the various $M_{\rm vir}$ bins. The intermediate phase metallicities are a bit higher than in the hot gas, within a factor 2, perhaps reflecting more efficient cooling of gas that enters the intermediate phase. For simplicity we set $\langle Z^\prime \rangle^{\rm it} = \langle Z^\prime \rangle^{\rm hot}$ as given by the PLM.

In Eq.~\ref{eq:Nionit} we assume that for any $M_{\rm vir}$ collisional ionization dominates at all intermediate phase temperatures, from $T_{\rm cool}=3\times 10^4$~K to $0.4T_{\rm vir}$, so that the mean CIE fraction for a given ion is 
\begin{equation} \label{eq:f_it_ion_CIE}
    \langle f_{\rm ion,CIE}\rangle^{\rm it} \equiv \frac{\int_{3 \times 10^{4} ~\rm{K}}^{0.4 T_{\rm vir}} f_{\rm ion,CIE}(T) dT}{0.4 T_{\rm vir} - 3 \times 10^{4} ~\rm{K}} ~~,
\end{equation}
where $f_{\rm ion,CIE}$ are the \cite{Gnat2007_CIE} CIE fractions shown in Fig.~\ref{fig:f_ion_CIE}.
In Fig. \ref{fig:IT_components} (middle panel) we plot the mean CIE fractions versus virial mass for our three ions O VI, O VII, and O VIII. The mean ion fractions are maximal when $0.4 T_{\rm vir}$ is close to the peak CIE temperatures for each ion. They decrease with higher $M_{\rm vir}$ and $T_{\rm vir}$ because the integrations are over a larger temperature range in which the CIE fractions are declining.

In Fig.~\ref{fig:IT_components} (right panel) we plot $\langle N_{\rm ion} \rangle^{\rm it}$ computed using Eq.~\ref{eq:M_ion} directly for intermediate temperature gas (solid curves), and using the approximation Eq.~\ref{eq:Nionit} (dashed curves). For each ion the two methods are in good agreement at sufficiently large $M_{\rm vir}$ at which collisions dominate the ionization. At low virial masses the intermediate phase is photoionized, as captured by the direct CI+PI computations (solid curves) but not by Eq.~\ref{eq:Nionit} (dashed curves) that includes only CIE by assumption. 

In the bottom row of Fig.~\ref{fig:plm_v_tng} we have added $\langle N_{\rm ion} \rangle^{\rm it}$ to the mean columns as estimated by the hot gas PLM as shown by the black dashed curve. This addition gives the red dashed curve. It is apparent that the collisionally ionized intermediate phase fully accounts for the mean ion columns at high virial masses. The photoionized intermediate temperature gas is negligible compared to the hot photoionized components that dominate at low masses. The intermediate phase starts contributing significantly to the mean columns, and eventually fully dominates, at $M_{\rm vir} \gtrsim 10^{12} ~\msun$, for O VI,  at $M_{\rm vir} \gtrsim  10^{13.4} ~\msun$, for O VII, and $M_{\rm vir} \gtrsim 10^{14} ~\msun$ for O VIII.

\section{Phase Diagrams} \label{sec:ions_different_mvirs}

Further insights into the origin of the CGM oxygen ions may be obtained with the aid of density-temperature $(n_H-T)$ phase diagrams for individual halos. We focus on three representative halos from our TNG100 sample. These are (a) a low mass halo with $M_{\rm vir}=1.3\times 10^{11}$~M$_\odot$ containing a central star-forming galaxy; (b) an intermediate mass halo at $1.2\times 10^{12}$~M$_\odot$, representative of the Milky Way galaxy scale; and (c) a massive halo at $8.9\times 10^{12}$~M$_\odot$ with a quenched central galaxy. The PLM parameters for each of these halos are close to the medians within their mass bins, and the associated ion mass distributions within them are representative.

\begin{figure*}
        \makebox[\textwidth][c]{\includegraphics[width= \textwidth]{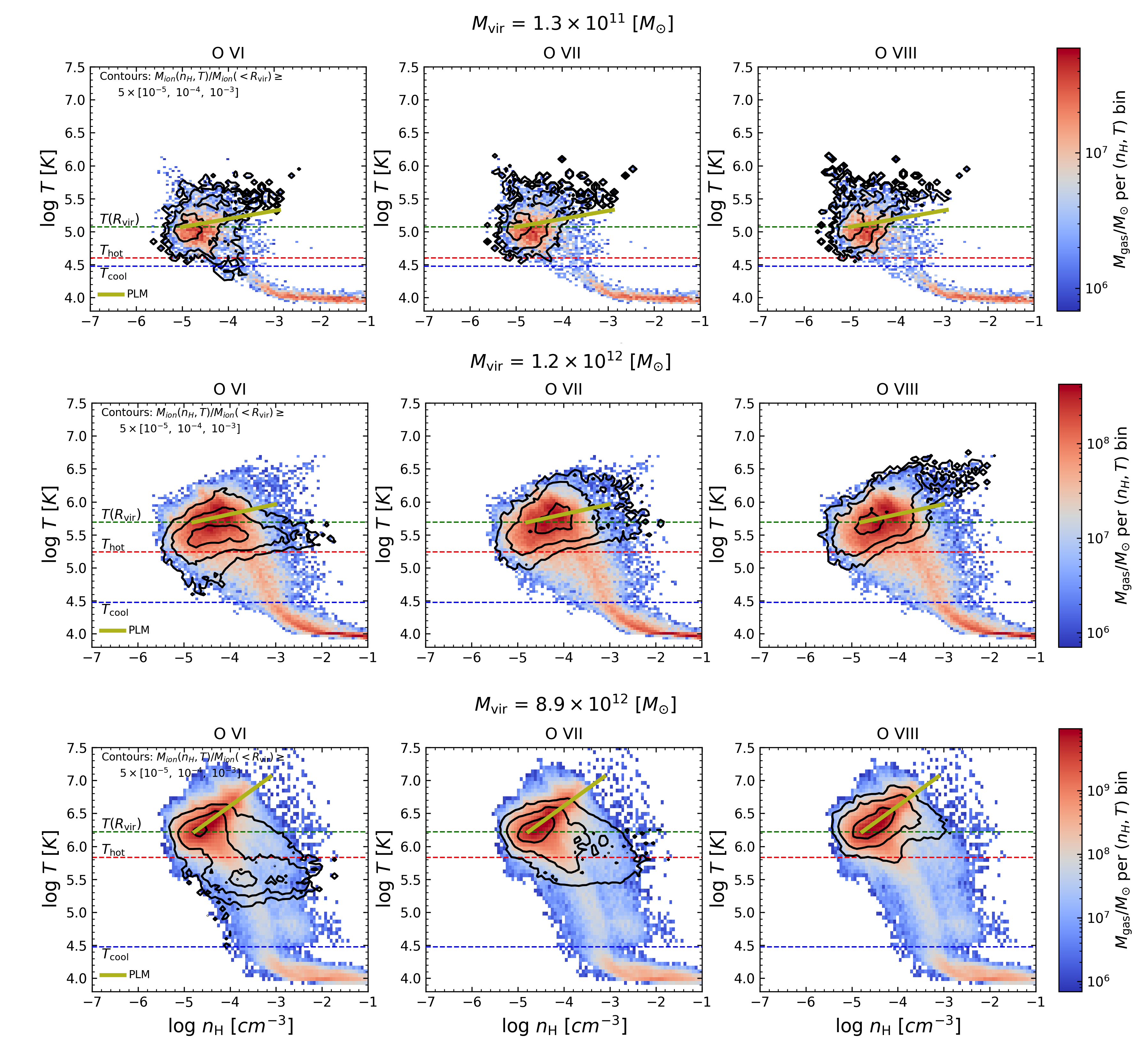}}
        \caption{Phase diagrams showing the distributions of total CGM gas mass (colored bins) and O VI (left), O VII (center) and O VIII (right) mass fractions (black contours) in the $n_H$ versus $T$ plane for our three representative halos. Top row: low-mass halo with $M_{\rm vir}=1.3\times 10^{11}$~M$_\odot$. Middle row: MW-mass halo with $M_{\rm vir}=1.2\times 10^{12}$~M$_\odot$. Bottom row: massive halo with $M_{\rm vir}=8.9\times 10^{12}$~M$_\odot$.
        The color scale shows the total gas mass per $(n_H, T)$ bin, where a $\log (n_H / {\rm cm^{-3}})$ bin is $0.08$ dex and a $\log (T / {\rm K})$ bin is $0.05$ dex.
        The black contours show where the fraction of the mass of a given ion per $(n_H,T)$ bin, relative to the total ion mass within $R_{\rm vir}$, are equal to $5\times 10^{-5}$, $5\times 10^{-4}$, and $5\times 10^{-3}$. The three horizontal lines in each panel mark (a) the gas temperature at the virial radius, $T(R_{\rm vir}) = \phi_T T_{\rm vir}$ (green), (b) the dividing line (red) at $0.4 T_{\rm vir}$ between the hot and intermediate phases, and (c) the dividing line (blue) at $3\times 10^{4}$ between intermediate and cool. The yellow line is our PLM for $T(n_H)$, from the outer to inner CGM, for this halo (see text).}
        \label{fig:phases}
    \end{figure*}

Our phase diagrams are presented in Fig.~\ref{fig:phases}, 
where each row corresponds to a separate halo, low mass (top), MW mass (middle), and high mass (bottom). In each panel the background blue to red points show the total gas mass per $(n_H, T)$ bin. Three horizontal lines are drawn indicating (in green) the mean temperature, $T(R_{\rm vir})=\phi_T T_{\rm vir}$ at the virial radius, the boundary (in red) at $0.4T_{\rm vir}$ separating the hot and intermediate temperature phases, and at $3\times 10^4$~K (in blue) marking the onset of the cool phase. The slanted solid yellow lines are the functions $T(n_H)$ as given by the PLMs for each halo. The PLM lines start at $n_H(R_{\rm vir})$ and $T(R_{\rm vir})$ and span the entire CGM to the inner denser and warmer points near the central galaxies. By definition the slopes of the PLM lines are $d{\rm log}(T)/d{\rm log}(n_H)=a_T/a_n$, where $a_T$ are $a_n$ are the temperature and density PLM indices.

For each halo we show three panels, one for each ion, with black contours showing the ion mass fractions relative to the total ion mass. The fractions are computed within each $n_H,T$ bin, and the displayed contours range from $5\times 10^{-5}$ to $5\times 10^{-3}$. The ion abundances are computed assuming the combination of collisional and photoionization (CI+PI). The contours show where in the $n_H$ versus $T$ plane the ion masses are built up, within the CGM volumes as a whole.

In Table \ref{table_mean_columns} we summarize the parameters and properties of our three illustrative halos. For each halo we list the virial mass, virial temperature, stellar mass, and specific star formation rate, and the TNG100 mean oxygen ion column densities assuming CI+PI, or CIE only. We also list the hot gas PLM parameters and the CI+PI mean ion columns computed using the PLMs and the intermediate gas corrections.


\begin{table}
\centering
\caption{Galaxy and halo properties, TNG100 mean column densities (for CI+PI and for CIE alone), PLM parameters, and PLM predictions for the mean column densities (for CI+PI) of the the low-mass, MW-mass, and massive fiducial halos analyzed in sections \ref{subsec:phase_1e11}, \ref{subsec:phase_MW_like}, and \ref{subsec:phase_1e13}.}
\label{table_mean_columns}
\begin{tabular}{@{}lccc@{}}
\toprule
\textbf{}                                                                & \textbf{Low-mass}      & \textbf{MW-mass}       & \textbf{Massive}      \\ \midrule
\multicolumn{1}{l|}{TNG100 Halo   Index}                                 & 9065                   & 1746                   & 171                    \\
\multicolumn{1}{l|}{$M_{\rm vir}   ~[\msun]$}                            & $1.3 \times 10^{11}$   & $1.2 \times 10^{11}$   & $8.9 \times 10^{12}$   \\
\multicolumn{1}{l|}{$T_{\rm vir}   ~[\rm{K}]$}                           & $1.0 \times 10^{5}$    & $4.3 \times 10^{5}$    & $1.7 \times 10^{6}$    \\
\multicolumn{1}{l|}{$M_{\rm *} ~ [\msun]$}                               & $1.1 \times 10^{8}$    & $2.1 \times 10^{10}$   & $6.7 \times 10^{10}$   \\
\multicolumn{1}{l|}{$SFR ~ [\msun \rm ~yr^{-1}]$}                        & $1.4 \times 10^{-2}$   & $1.7$                  & $1.4 \times 10^{-2}$   \\
\multicolumn{1}{l|}{$sSFR ~ [\rm{yr^{-1}]}$}                             & $1.4 \times 10^{-10}$  & $7.8 \times 10^{-11}$  & $2.2 \times 10^{-13}$ \\ 
\multicolumn{1}{l|}{$f_{\rm cCGM}$}                             & $0.16$  & $0.10$  & $0.03$ \\ 
\multicolumn{1}{l|}{$f_{\rm itCGM}$}                             & $0.01$  & $0.08$  & $0.02$ \\ 
\multicolumn{1}{l|}{$f_{\rm hCGM}$}                             & $0.33$  & $0.46$  & $0.46$ \\  \midrule
\multicolumn{4}{c}{\textbf{TNG100 mean columns   (CI+PI)}}                                                                                          \\ \midrule
\multicolumn{1}{l|}{$\langle   N_{\rm O~VI} \rangle ~ [\rm{cm^{-2}}]$}   & $9.4 \times 10^{12}$   & $4.0 \times 10^{14}$   & $2.2 \times 10^{13}$   \\
\multicolumn{1}{l|}{$\langle   N_{\rm O~VII} \rangle ~ [\rm{cm^{-2}}]$}  & $2.6 \times 10^{13}$   & $3.8 \times 10^{15}$   & $2.1 \times 10^{15}$   \\
\multicolumn{1}{l|}{$\langle   N_{\rm O~VIII} \rangle ~ [\rm{cm^{-2}}]$} & $2.1 \times 10^{12} $  & $3.9 \times 10^{14} $  & $1.8 \times 10^{15} $  \\ \midrule
\multicolumn{4}{c}{\textbf{TNG100 mean columns   (CIE)}}                                                                                            \\ \midrule
\multicolumn{1}{l|}{$\langle   N_{\rm O~VI} \rangle ~ [\rm{cm^{-2}}]$}   & $3.3 \times 10^{12}$   & $3.4 \times 10^{14}$   & $1.8 \times 10^{13}$   \\
\multicolumn{1}{l|}{$\langle   N_{\rm O~VII} \rangle ~ [\rm{cm^{-2}}]$}  & $1.2 \times 10^{13}$   & $3.9 \times 10^{15}$   & $2.7 \times 10^{15}$   \\
\multicolumn{1}{l|}{$\langle   N_{\rm O~VIII} \rangle ~ [\rm{cm^{-2}}]$} & $3.5 \times 10^{9} $   & $7.2 \times 10^{13} $  & $1.4 \times 10^{15}$   \\ \midrule
\multicolumn{4}{c}{\textbf{PLM parameters}}                                                                                                         \\ \midrule
\multicolumn{1}{l|}{$a_{\rm P,   th}$}                                   & $1.69$                 & $1.47$                 & $1.78$                 \\
\multicolumn{1}{l|}{$a_{\rm n}$}                                         & $1.51$                 & $1.27$                 & $1.18$                 \\
\multicolumn{1}{l|}{$a_{\rm Z}$}                                         & $0.81$                 & $0.65$                 & $0.37$                 \\
\multicolumn{1}{l|}{$\phi_T$}                                            & $1.18$                 & $1.13$                 & $0.97$                 \\
\multicolumn{1}{l|}{$f_{\rm hCGM}$}                                      & $0.33$                 & $0.46$                 & $0.46$                 \\
\multicolumn{1}{l|}{$Z^{\prime}_{\rm   vir}$}                            & $0.01$                 & $0.18$                 & $0.16$                 \\ \midrule
\multicolumn{4}{c}{\textbf{PLM mean columns   (CI+PI)}}                                                                                             \\ \midrule
\multicolumn{1}{l|}{$\langle   N_{\rm O~VI} \rangle ~ [\rm{cm^{-2}}]$}   & $5.8 \times 10^{12}$   & $1.2 \times 10^{14}$   & $5.6 \times 10^{12}$   \\
\multicolumn{1}{l|}{$\langle   N_{\rm O~VII} \rangle ~ [\rm{cm^{-2}}]$}  & $1.3 \times 10^{13}$   & $3.9 \times 10^{15}$   & $1.9 \times 10^{15}$   \\
\multicolumn{1}{l|}{$\langle   N_{\rm O~VIII} \rangle ~ [\rm{cm^{-2}}]$} & $9.5 \times 10^{11} $  & $4.1 \times 10^{14} $  & $2.3 \times 10^{15} $  \\ \bottomrule
\end{tabular}
\end{table}

\subsection{Low-mass halo}\label{subsec:phase_1e11}
Our low mass halo is for $M_{\rm vir}=1.3\times 10^{11}$~M$_{\odot}$. 
Its central galaxy is star forming, with a stellar mass $M_*=1.0\times 10^8$~M$_{\odot}$ and a star formation rate $SFR=1.38\times 10^{-2}$~M$_\odot$~yr$^{-1}$. The total CGM gas fraction in all phases, hot, intermediate, and cold is $f_{\rm CGM}=0.50$, corresponding to a total CGM gas mass of $9.9 \times 10^{9} ~\msun$.

The phase diagrams for this halo in Fig.~\ref{fig:phases} show that the CGM gas mass is concentrated in two regions. First is a narrow strip at $T\sim 10^4$~K with $n_H\gtrsim 10^{-3}$~cm$^{-3}$. This is cool gas with a small volume filling factor of $7.1 \times 10^{-3}$, that is likely falling towards the galaxy and feeding star-formation. The gas mass faction in the cool phase is $0.16$. Second is around $T\approx T_{\rm vir} = 1.0\times 10^5$~K and $n_H\sim 10^{-5}$~cm$^{-3}$. This is hot phase volume filling gas, with a CGM mass fraction $0.33$. Because the halo mass is low, with a correspondingly low $T_{\rm vir}$, the intermediate temperature component lies within the narrow range $4.5 \lesssim \log(T/K) \lesssim 4.6$, and with a mass fraction equal to $0.01$.

The black contours in left panel of Fig.~\ref{fig:phases} (top row) show that the O VI arises in two zones, around $T \approx 10^{5}$ K and $n_H \approx 10^{-5} ~\rm{cm^{-3}}$, and around $T \approx 3 \times 10^{5}$ K and $n_H \approx 10^{-4} ~\rm{cm^{-3}}$. These are both in the hot phase, but the first region contains collisionally ionized O VI, whereas the second is photoionized O VI (see also Figure \ref{fig:f_ion_full}). Most of the O VI is in the photoionized component, consistent with our conclusion in \S~\ref{sec:plm} that photoionization dominates when $T_{\rm vir}$ is below the CIE peak temperature.

The behavior for O VII and O VIII is similar, again because $T_{\rm vir}$ is below the CIE peak temperatures for both ions. These ions are both produced in the hot phase, but this gas is not hot enough for efficient collisional ionization, except for a few gas particles with large temperature fluctuations. Most of the O VII and O VIII is photoionized, consistent with our findings in \S~\ref{sec:plm}. Indeed, Fig.~\ref{fig:plm_v_tng} shows that at $1.3\times 10^{11}$~M$_\odot$, the mean column \Noviii is larger for CI+PI compared to CIE, by more than a factor of 2.

For O VIII, and when considering CIE alone, \foviii$ \approx 0.01$ at the highest CGM gas temperatures of $T \approx 10^6$ K and drops significantly as $T$ decreases. In practice, only a very small fraction of the CGM participates in generating O VIII under CIE. However, when photoionization by the MGRF is included as well, O VIII is generated over a much larger temperature range, going down to $T_{\rm hot} = 0.4 T_{\rm vir}$. In addition to a larger fraction of the CGM participating in generating O VIII, the associated ion fractions are significantly higher, reaching \foviii$ \approx 0.3$. Consequently, the total O VIII mean column under CI+PI is larger by a factor of $\sim 600$ compared to under CIE alone.
 
Since all three oxygen ions originate in the hot phase for this halo mass, our PLM together with CI+PI ionization provides an accurate estimate for the mean columns. The corrections for the intermediate phase are unnecessary.

We stress that although photoionization dominates the ion production, it is not setting the temperature. The hot phase gas temperature is much higher than the $\sim 10^{4}$ K we would expect for gas that is in equilibrium with the MGRF. This gas is shock heated as it falls onto the halo, and the MGRF primarily determines the oxygen ionization states, but not the heating.

\subsection{MW-mass halo} \label{subsec:phase_MW_like}
Next is a TNG100 halo with $M_{\rm vir}=1.2 \times 10^{12} ~\rm{M_{\odot}}$, with a central star forming galaxy with stellar mass $M_*=2.1 \times 10^{10} ~ \rm{M_{\odot}}$. The star formation rate is 1.64~M$_\odot$~yr$^{-1}$. This halo is representative of Milky Way mass galaxies in TNG100\footnote{While this galaxy mass is typical for a MW-mass halo in TNG100, the actual Milky Way galaxy is quite different: while \cite{Posti19_MW} suggest the MW virial mass to be $1.3 \pm 0.3 \times 10^{12} ~\msun$, \cite{Licquia2016_MW_M_star} find its stellar mass to be $5.7^{+ 1.5}_{- 1.1} \times 10^{10} ~\msun$, a factor of $\sim 2.7$ higher than the median TNG100 galaxy with a similar halo mass.}. The total CGM gas fraction in all phases is $f_{\rm CGM}=0.64$ for a total CGM gas mass of $1.2 \times 10^{11} ~\msun$.

The phase diagrams for this halo in Fig. \ref{fig:phases} again show a cool gas strip at $T\sim 10^4$~K at densities $n_H \gtrsim 10^{-3}$~cm$^{-3}$. The mass fraction in cool gas is $0.10$. At the virial temperature of $4.3\times 10^5$~K for this halo, the intermediate phase ranges from $3\times 10^4$~K to $1.7\times 10^5$~K. However, the mass fraction in the intermediate temperature gas remains small at $0.08$, suggesting that this phase is transitional, either cooling down or heating up. Most of the CGM gas mass is in the hot phase, close to the virial temperature, and with densities between $10^{-5}$ and $10^{-4}$~cm$^{-3}$. The mass fraction in this component is $0.46$.

All three oxygen ions are produced in the hot phase in this halo. Because the hot gas temperature is close to the CIE peak temperatures for O VI and O VII these ions are produced by collisional ionization. However, except for a few temperature fluctuations, the hot gas is still not hot enough to enable collisional ionization of the O VIII, and this ion is produced mainly by photoionization. Temperature fluctuations of up to $\sim 1.6\times 10^6$~K do occur at densities near $10^{-4}$~cm$^{-3}$, at $r\approx 0.6R_{\rm vir}$, for which collisional ionization is effective for the O VIII, but the contributions from such gas is small. The conclusion that O VIII is produced by photoionization in MW-mass halos is consistent with the CGM model presented by \citetalias{Faerman2020} (see their Fig.~6).

\subsection{Massive Halo} \label{subsec:phase_1e13}
Our third halo is massive, with $M_{\rm vir}=8.9 \times 10^{12} ~\rm{M_{\odot}}$, $M_*=6.6\times 10^{10}$~M$_\odot$, and an $SFR=1.4\times 10^{-2}$~M$_\odot$~yr$^{-1}$. The star formation rate is very low for this mass and the galaxy is quenched. The total CGM gas fraction is $f_{\rm CGM}=0.51$ for a total CGM gas mass of $7.1 \times 10^{11} ~\msun$.

The cool gas tail is still present in the phase plots for this halo, but the gas mass fraction in cool gas is reduced to $0.03$ \citep[see also Figure 8 in][]{Oren2024}. At the virial temperature of $1.7\times 10^6$~K for this halo, the intermediate phase ranges from $3\times 10^4$~K to $6.8\times 10^5$~K, and the gas fraction is $0.02$. Most of the CGM gas mass is in the hot phase, with particle temperatures from $\sim 10^6$ to $\sim 10^7$~K, and densities from $\sim 10^{-5}$ to $3\times 10^{-4}$~cm$^{-3}$, as also tracked by the PLM line. The hot gas fraction is $0.46$, corresponding to a gas mass of $6.5 \times 10^{11} ~\msun$.

The phase diagram shows that the O VI arises at two loci, in the intermediate phase near $3\times 10^5$~K, and in the hot phase near $2\times 10^6$~K. The gas fraction in the intermediate phase is small, but the temperature is near the O VI CIE peak, and a large O VI mass is produced. Conversely, the CIE O VI fraction in the hot gas is small ($6\times 10^{-3}$ at $2\times 10^6$~K) but the mass of hot gas is large, leading to a significant contribution to the O VI mass. Collisional ionization dominates at both locations, and photoionization is negligible as indicated by Fig.~\ref{fig:plm_v_tng}. 
For this halo 68\% of the O VI is produced in the intermediate phase, and 32\% in the hot phase (see Fig.~\ref{fig:mass_fractions}).

The O VII and O VIII are both produced in the hot phase, which for this halo mass is hot enough for collisional ionization to dominate, with CIE peaks for both ions close to the virial temperature. O VII is produced in CIE over a broad range of temperatures, from $3\times 10^{5}$ to $2\times 10^{6}$~K, (see Fig.~\ref{fig:f_ion_CIE}) and around 10\% of the O VII is produced collisionally within the intermediate phase. For O VIII this halo mass is near to the critical mass below which photoionization begins to contribute significantly to the ion production (as seen in Fig.~\ref{fig:plm_v_tng}).

\subsection{Cooling Time Scales}\label{Cooling_times}

It is of interest to consider the cooling time scales within the CGM gas phases, as traced by the oxygen ions. For this purpose we focus on O VI.

\begin{figure*}
        \makebox[\textwidth][c]{\includegraphics[width= \textwidth]{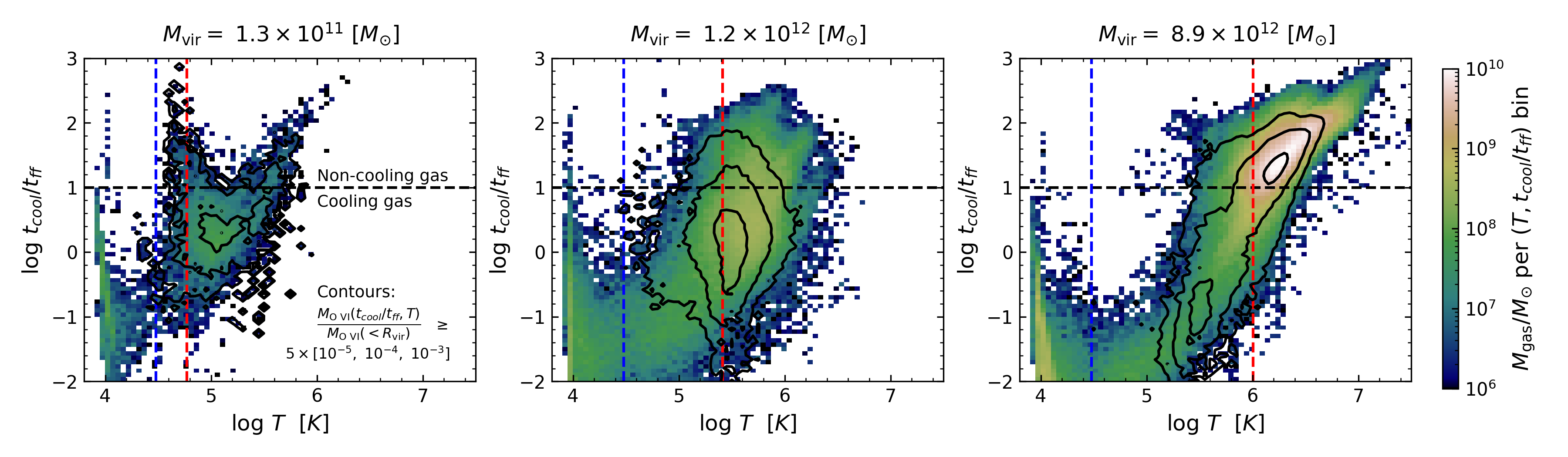}}
        \caption{Total CGM gas mass and O VI mass distributions within $R_{\rm vir}$ in the $T$ versus $t_{\rm cool} / t_{\rm ff}$ planes (see text) for our three representative halos (left, middle, right). The gas mass fractions are indicated by the colored bins. The color scale shows the total gas mass per $(T, t_{\rm cool} / t_{\rm ff})$ bin, where a $\log (T / {\rm K})$ bin is $0.05$ dex and a $\log (t_{\rm cool} / t_{\rm ff})$ bin is $0.07$ dex.
        The black contours show the O VI mass fractions at the $5\times 10^{-5}$, $5\times 10^{-4}$ and $5\times 10^{-3}$ levels. The vertical lines mark temperatures of $3\times 10^4$~K and $0.4T_{\rm vir}$ that separate the cool, intermediate and hot phases.}
    
        \label{fig:OVI_cooling_TNG}
    \end{figure*}

In Fig.~\ref{fig:OVI_cooling_TNG} we plot total gas mass distributions (color coded) together with the O VI distributions (black contours) in the $T$ versus $t_{\rm cool}/t_{\rm ff}$ planes for our three representative halos. The cooling and free-fall times are defined as
\begin{equation}
t_{\rm cool} = \frac{3}{2} \frac{n k_{\rm B}T} {n_{\rm e}n_{\rm H}\Lambda} \ \ , \ \ t_{\rm ff} = \sqrt{\frac{2r}{g}} \ \ \ ,
\end{equation}
where $\Lambda$ is the cooling function (erg cm$^{3}$ s$^{-1}$) as given by TNG100 for every gas particle, and $g$ is the gravitational acceleration at $r$ due to all mass components, gas, stars, and dark matter. The horizontal lines in Fig.~\ref{fig:OVI_cooling_TNG} are set at $t_{\rm cool}/t_{\rm ff}=10$, a characteristic value below which radiating gas may become thermally unstable to rapid cooling and filamentary collapse (\citealp{Sharma2012b,Voit2017_precip};\citetalias{Faerman2020}).

For the low-mass halo, and most of the MW-mass halo,  $t_{\rm cool}/t_{\rm ff} \lesssim 10$ suggesting that the CGM in these halos would be subjected to thermal instability in the absence of feedback and heating. 
As discussed above, for these halos most of the O VI is produced in gas close to the halo virial temperatures, via photoionization in the low-mass halo and collisional ionization in the MW-mass halo. The O VI is clearly formed in the unstable regime, as indicated by the black contours that peak well below the horizontal instability line. For the MW-mass halo $t_{\rm cool}/t_{\rm ff} \lesssim 4$, consistent with the upper limit on this ratio derived by \citetalias{Faerman2020} (see their Eq.~26) for a $10^{12}$~M$_\odot$ halo with an O VI column density of $\sim 3\times 10^{14}$~cm$^{-2}$. This is indeed the mean TNG100 O VI column density at this halo mass scale, as shown in Fig.~\ref{fig:plm_v_tng}.

For our massive halo (right panel) most of the gas is in the stable regime, where the cooling times are very long due to the high temperatures and inefficient cooling in the hot phase for this mass. However, the intermediate temperature gas lies below the instability line. As discussed above, for this halo the O VI is formed by collisional ionization in both the hot phase, in which the O VI fractions are small but the overall gas mass is large, and in the intermediate phase at temperatures near the CIE peaks where the O VI fractions are large, but with much lower gas mass. This behavior is reflected in Figure~\ref{fig:OVI_cooling_TNG} which shows the two O VI concentrations, one in stable hot gas with a long cooling time, and the second in unstable intermediate gas with a short cooling time. Most of the O VI (75\%) is actually formed in the unstable intermediate temperature regime.

Is O VI a tracer of thermally unstable gas? For MW-mass and lower mass halos, cooling gas is unstable since $t_{\rm cool}/t_{\rm ff} \lesssim 10$. However, in the presence of feedback, some O VI is formed in gas that is heating up and may be stable. In more massive halos the the O VI may be formed and cooling in both the stable and unstable regimes.

\section{Summary}\label{sec:summary}
In this paper we investigate the origin of the highly ionized oxygen species, O VI, O VII, and O VIII, in the circumgalactic medium (CGM) of galaxies at redshifts $z\approx 0$, using the outputs of the Illustris TNG100 simulations, in combination with post-processing calculations of the ionization states, together with analytical modeling based on the methodology presented in \cite{Oren2024}. The work presented here is complementary to the phenomenological analysis presented by \cite{Faerman2017} and \cite{Faerman2020}, and the Illustris TNG oxygen study by \cite{Nelson2018_TNG_absorption}. 

As we discuss in \S~\ref{sec:ionization}, our focus is on the competition between electron impact collisional ionization (CI), and photoionization (PI) by background metagalactic UV/Xray radiation. We use the {\it Cloudy} photoionization code \citep{Ferland2017_CLOUDY} to compute local oxygen ionization fractions as functions of gas temperature and density for optically thin conditions, and the fractions are then assigned to every CGM gas particle in the TNG100 halo outputs. In combination with the TNG100 oxygen abundances we are then able to construct CGM ion column density maps for a wide range of halo virial masses. Our TNG100 halo sample is described in 
\S~\ref{sec:illustris}. It consists of a total of 785 halos, 502 with central star-forming galaxies, and 283 containing quenched galaxies. The virial mass range is $10^{10.5}$ to $10^{14.5}$~M$_\odot$.

We describe our column density maps in \S~\ref{sec:oxymaps}. We construct two families of maps. First are ``halo cuts'' that include oxygen ions contained within the virial radii only. Second are ``2Mpc cuts'' that include gas particles out to about 2~Mpc from the central galaxies, well beyond the virial radii. 
The two map families enable us to separate out background intergalactic contributions to the integrated oxygen column densities (see also \citealp{Bromberg2025_IGM_OVI}). We use our maps to generate radial column density profiles, for direct comparisons to available absorption and emission line spectroscopic observations. We find that computations including both collisional and photoionization (CI+PI) predict oxygen ion masses and column densities that are broadly consistent with observation across a wide range of halo masses.

Following \cite{Oren2024} we divide the CGM into three temperature regimes, or gas phases. First, is hot shock heated virialized gas with $T>0.4 T_{\rm vir}$. This temperature cut is based on our finding in \cite{Oren2024} that more than 90\% of the thermal Sunyaev-Zeldovich (tSZ) signals produced in TNG100 halos arise from gas above this critical temperature. Second, is cool gas at temperatures below $3\times 10^4$~K, characteristic of gas heated purely by photoionization. Third, is intermediate temperature gas between hot and cool. 
In \S~\ref{sec:plm} we present results for the volume filling factors of these three gas phases, hot, intermediate, and cool, as functions of virial mass across our halo sample. 

To interpret our computations we make use of the simple analytic radial power-law models (PLMs) that we presented in \cite{Oren2024} for the density and temperature distributions of the hot CGM gas in the TNG100 halos. We review our PLM in \S~\ref{sec:plm}, and extend it to include power-law fits for the radially dependent metallicity distributions in the TNG100 halos. We use the PLM parameters to analytically estimate the ionization states, distributions, and integrated column densities of the oxygen ions produced in the hot components of the TNG100 halos.

In \S~\ref{sec:mean_columns} we present computations for the mean oxygen ion column densities produced within the TNG100 halos as functions of virial mass. The mean columns may be used to predict observable absorption line strengths. To determine the relative roles of collisional ionization versus photoionization in producing the oxygen ions, and to identify the gas phases within which the ions reside, we compute  the mean columns in four ways. First, assuming CI+PI applied to the TNG100 particle data outputs, second assuming just CIE for the TNG100 outputs, third assuming CI+PI for the hot gas PLM representations, and fourth assuming just CIE for an additional analytic model for the intermediate temperature phase. We also provide convenient fitting formulae for the mean oxygen column densities as functions of halo virial mass.

We find, as a general rule, that if the halo virial temperature is below the gas temperature at which the ion fraction is maximal for CIE, then photoionization contributes significantly or dominates the overall CGM ion mass and column density. However, when photoionization does dominate, the ions are produced primarily in the hot phase near the virial temperature of the given halo (and not in cool gas). Conversely, if the virial temperature is above the gas temperatures at which the ion fractions are maximal for CIE, then collisions dominate the ionization. However, if the virial temperature is much higher than the ion CIE peak temperature then a substantial or dominating portion of the ion mass may reside within the intermediate temperature phase as opposed to the hot virialized gas.

Thus, a key conclusion of this work is that the dominant ionization mechanism of the high oxygen ions and the thermal state of the gas should not be conflated. The ionization state is governed primarily by the competition between collisional ionization and photoionization, whereas the gas temperature reflects the thermal history of the gas. Throughout much of the halo mass range considered here, we find that photoionized high oxygen ions are formed in gas that remains shock-heated to approximately the virial temperature, rather than in gas thermally equilibrated with the metagalactic radiation field. Consequently, there is generally no regime in which both the high oxygen ionization and the heating are simultaneously controlled by photoionization. This contrasts with the common interpretation that any photoionized gas necessarily resides at temperatures of order $10^4$~K. 

To further elucidate the origin of the oxygen ions and the dependence on halo mass, in \S~\ref{sec:ions_different_mvirs} we show the total gas and oxygen ion mass distributions across density versus temperature phase diagrams, for three representative halos. We present results for a low mass halo at $\sim 10^{11}$~M$_\odot$, for a halo at the Milky-Way (MW) mass scale $\sim 10^{12}$~M$_\odot$, and for a high mass halo at $\sim 10^{13}$~M$_\odot$. The phase diagrams include gas in all three thermal phases, within the entire CGM volumes. For the low mass halo, the virial temperature is low, and all three ions are produced mainly by photoionization in the hot phase, with a small contribution to the O VI via collisional ionization. For the MW-mass halo, all three ions are also produced in the hot virialized gas, and the O VI and O VII are collisionally ionized. However, the O VIII remains photoionized. In the massive halo all three ions are collisionally ionized, and photoionization is negligible.
Most of the O VI is produced in the intermediate temperature phase rather than in the hot gas. This is because for this halo mass the intermediate phase brackets the CIE peak temperatures at which the OVI is most efficiently produced.  A portion of the O VI is nevertheless produced in the hot gas because of the much larger overall gas mass in this phase. We point out
that for the massive halo the intermediate phase may be thermally unstable to rapid cooling, as opposed to the hot virialized gas, which is likely stable.

In this work we focus on the origin of the highly ionized oxygen species, O VI, O VII, and O VIII, in the circumgalactic medium of galaxies. But other high ionization CGM tracers are available and have been observed, such as the isoelectronic ions C IV \citep{Chen2001,Garza2025}, N V \citep{Werk2016}, and Ne VIII \citep{Burchett2019,Qu2024}.
Our methodology is general and may be applied to these and other elements and ionization stages in the CGM, for determinations of their origin, over a range of halo masses and redshifts.



\section*{Acknowledgments}
We thank Shy Genel, Amit Nestor-Shachar, and Jonathan Stern for helpful discussions. This work was supported by the German Science Foundation via DFG/DIP grant STE/ 1869-2 GE/ 625 17-1, by the Center for Computational Astrophysics (CCA) of the Flatiron Institute, and by the Mathematical and Physical Sciences (MPS) division of the Simons Foundation, USA. The research of CFM was supported in part by the NASA ATP grant 80NSSC20K0530 and in part by grant NSF PHY2309135 to the Kavli Institute for Theoretical Physics (KITP). YF is supported by the Tel Aviv University - Harvard Astronomy Program.




\bibliographystyle{mnras}
\bibliography{absorption_tng} 



\appendix

\section{Absorption maps of selected halos}\label{appendix:maps}
\begin{figure*}
        \makebox[\textwidth][c]{\includegraphics[width= \textwidth] {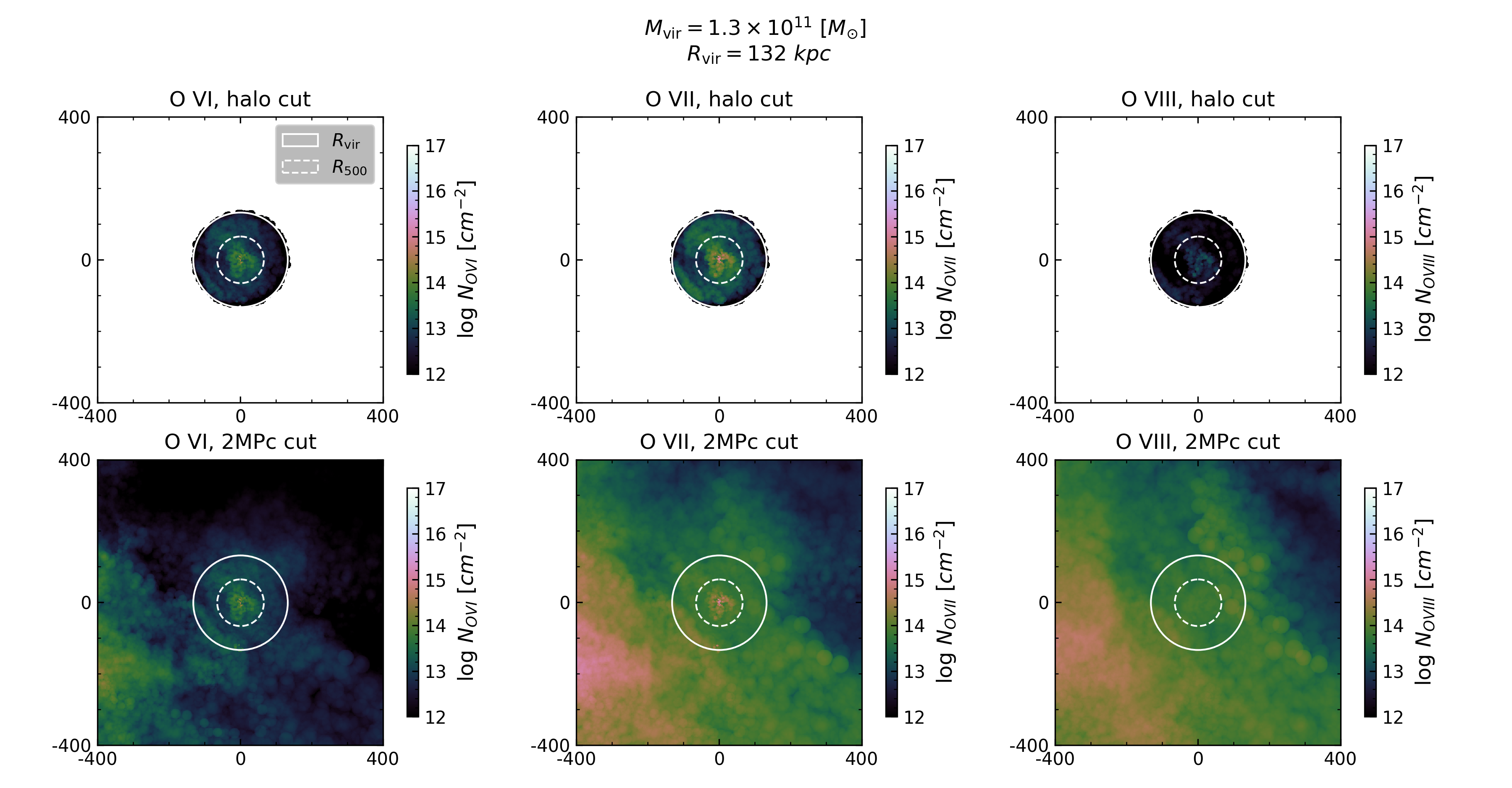}}
        \caption{O VI (left), O VII (center), and O VIII (right) column density maps for our representative low-mass halo with $M_{\rm vir}=1.3\times 10^{11}$~M$_\odot$ as described in \S~\ref{subsec:n_ion_maps}. Top row are the halo cut maps, and bottom row are the 2Mpc cut maps.}
        \label{fig:maps_1e11}
    \end{figure*}

\begin{figure*}
        \makebox[\textwidth][c]{\includegraphics[width= \textwidth] {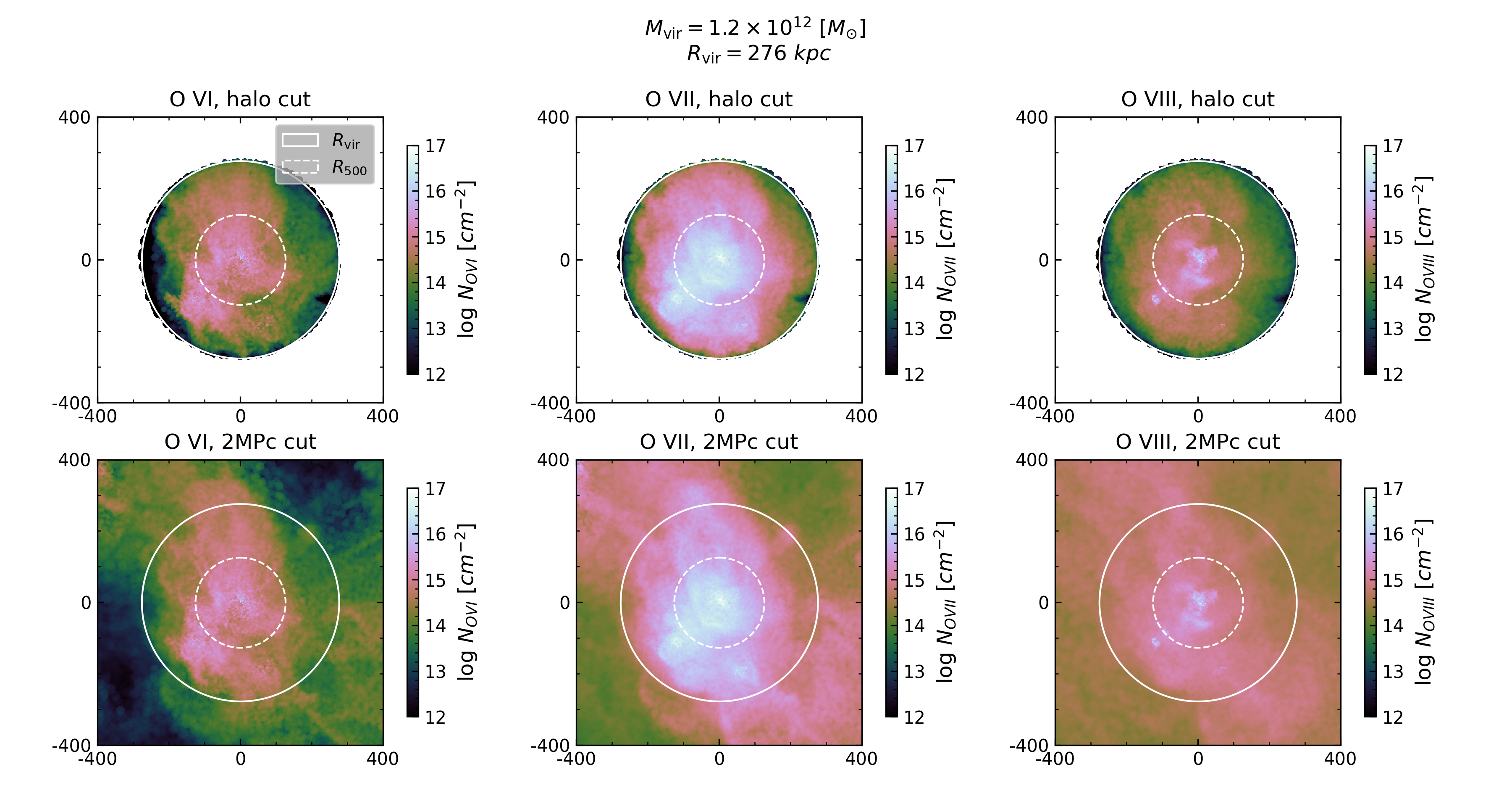}}
        \caption{Same as Fig. \ref{fig:maps_1e11} but for $M_{\rm vir}=1.2\times 10^{12}$~M$_\odot$.}
        \label{fig:maps_1e12}
    \end{figure*}

\begin{figure*}
        \makebox[\textwidth][c]{\includegraphics[width= \textwidth] {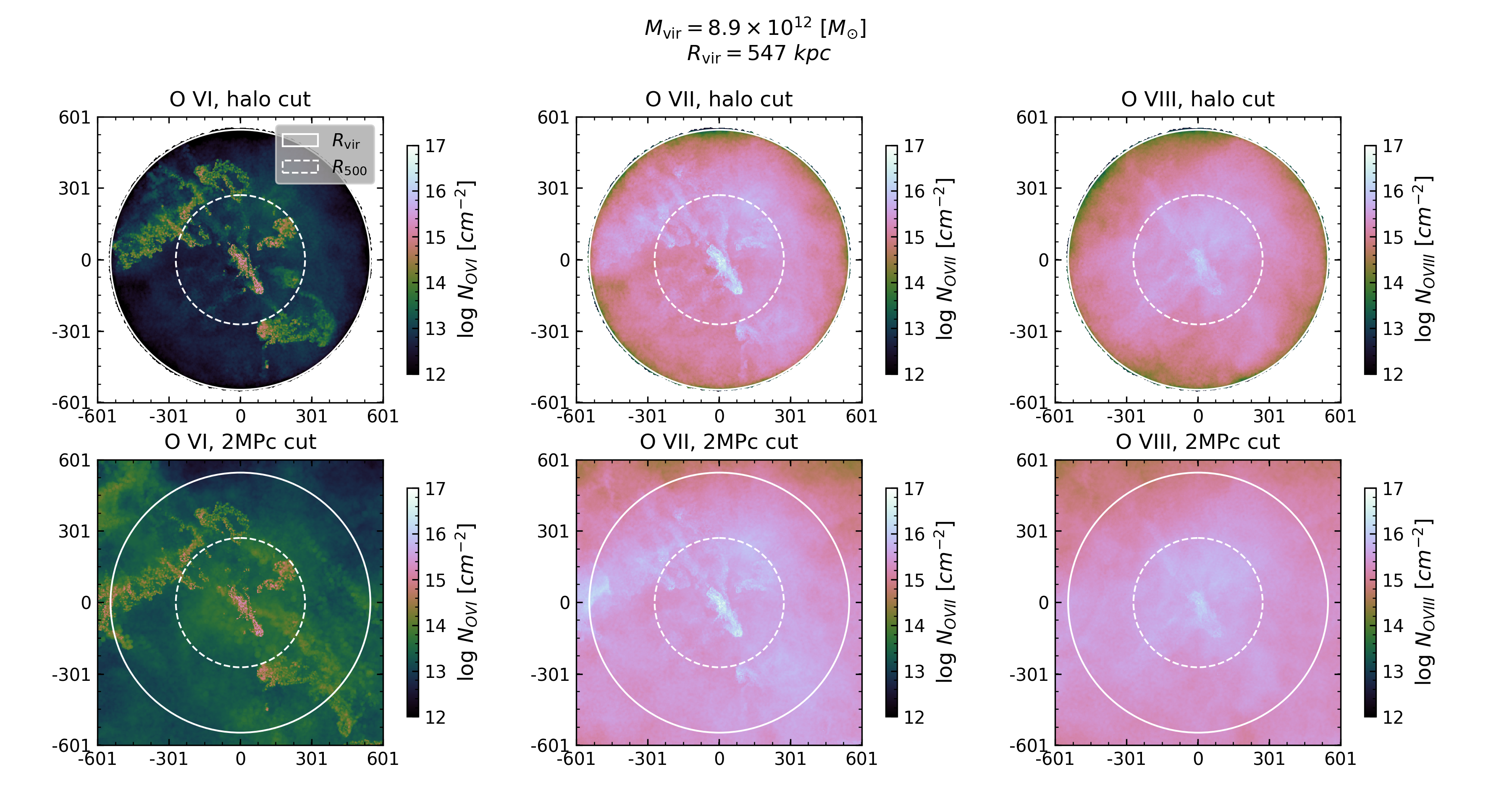}}
        \caption{Same as Fig. \ref{fig:maps_1e11} but for $M_{\rm vir}=8.9\times 10^{12}$~M$_\odot$.}
        \label{fig:maps_1e13}
    \end{figure*}

In Figs. \ref{fig:maps_1e11}, \ref{fig:maps_1e12}, and \ref{fig:maps_1e13} we plot the O VI, O VII, and O VIII column density maps for our three fiducial TNG100 halos discussed in \S~\ref{sec:ions_different_mvirs}.  We show maps from both the halo cut (top row) and the 2Mpc cut (bottom row). Each map is bordered at the larger between 400 kpc and $1.1 R_{\rm vir}$, where the latter is only applied in Fig. \ref{fig:maps_1e13} as the virial radius of this massive halo is 547 kpc. 

\bsp	
\label{lastpage}
\end{document}